\documentclass{emulateapj}

\usepackage{color}
\definecolor{dr}{rgb}{0.6,0,0}
\definecolor{db}{rgb}{0,0,0.6}

\usepackage[ hyperfootnotes={false},
dvips,
pdftitle={Stellar mass dependent disk dispersal},
pdfauthor={Grant Kennedy and Scott Kenyon},
pdfstartview={FitH},
bookmarksopen={true} ]{hyperref}
\hypersetup{colorlinks={true}, urlcolor={db}, citecolor={dr}}

\newcommand{\ha}{EW[H$\alpha$]}

\begin{document}

\title{Stellar mass dependent disk dispersal}

\slugcomment{Accepted to ApJ 17 Jan 2009}

\author{Grant M. Kennedy}
\affil{Research School of Astronomy and Astrophysics, Australian National University, Canberra, Australia}
\email{grant@mso.anu.edu.au}

\and

\author{Scott J. Kenyon}
\affil{Smithsonian Astrophysical Observatory, Cambridge, MA 02138, USA}
\email{kenyon@cfa.harvard.edu}

\shortauthors{KENNEDY \& KENYON}
\shorttitle{STELLAR MASS DEPENDENT DISK DISPERSAL}

\begin{abstract}
  We use published optical spectral and IR excess data from nine young
  clusters and associations to study the stellar mass dependent
  dispersal of circumstellar disks. All clusters older than
  $\sim$3\,Myr show a decrease in disk fraction with increasing
  stellar mass for Solar to higher mass stars. This result is
  significant at about the 1\,$\sigma$ level in each cluster. For the
  complete set of clusters we reject the null hypothesis---that Solar
  and intermediate-mass stars lose their disks at the same rate---with
  95--99.9\% confidence. To interpret this behaviour, we investigate
  the impact of grain growth, binary companions, and photoevaporation
  on the evolution of disk signatures. Changes in grain growth
  timescales at fixed disk temperature may explain why early-type
  stars with IR excesses appear to evolve faster than their later-type
  counterparts. Little evidence that binary companions affect disk
  evolution suggests that photoevaporation is the more likely
  mechanism for disk dispersal. A simple photoevaporation model
  provides a good fit to the observed disk fractions for Solar and
  intermediate-mass stars. Although the current mass-dependent disk
  dispersal signal is not strong, larger and more complete samples of
  clusters with ages of 3--5\,Myr can improve the significance and
  provide better tests of theoretical models. In addition, the orbits
  of extra-Solar planets can constrain models of disk dispersal and
  migration. We suggest that the signature of stellar mass dependent
  disk dispersal due to photoevaporation may be present in the orbits
  of observed extra Solar planets. Planets orbiting hosts more massive
  than $\sim$1.6\,$M_\odot$ may have larger orbits because the disks
  in which they formed were dispersed before they could migrate.
\end{abstract}


\section{Introduction}

Most known extra-Solar planets orbit roughly Solar-mass stars; the
result of observational biases towards these stars in planet hunting
surveys. Recently, planets orbiting both low and intermediate mass
stars have been discovered
\citep[e.g.][]{2005ApJ...634..625R,2007ApJ...665..785J}, thus
increasing the diversity of planet host stars. In parallel, there have
been various models proposed that attempt to explain and predict the
frequency and properties of these planets as a function of stellar
mass
\citep{2005ApJ...626.1045I,2006A&A...458..661K,2007ApJ...660..845B,2008ApJ...673..502K}.

These models need observational constraints, which are provided in two
ways. The observed properties of the planets yields one set of
constraints, setting the final distributions that models must
reproduce. These distributions contain trends such as an increasing
planet frequency with increasing stellar metallicity
\citep{2005ApJ...622.1102F}. However, because planets are observed
around main-sequence stars, these constraints provide no direct
information about the circumstellar environment during the 1-100\,Myr
epoch of planet formation.

Planets form in circumstellar disks, and observations of these disks
yield another set of constraints. These constraints are used as model
parameters. For example, disks are made up of gas and dust and in many
cases have enough material to form planetary systems with planets as
massive as those in our Solar System
\citep[e.g.][]{2000prpl.conf..559N,2005ApJ...631.1134A}. Of these
constraints, the strongest is the observed disk lifetime. The disk
dispersal timescale sets the time taken for dust to grow into
$\sim$Earth-mass objects and accrete gas to form giant planets.

The timescale for disk dispersal is inferred from dust and/or gas
signatures. The fraction of stars with disks in clusters of roughly
coeval young stars decreases with increasing age, yielding the typical
lifetime of circumstellar disks. The disk lifetime is usually derived
from the dust signature \citep[e.g.][]{2001ApJ...553L.153H}, with the
assumption that the gaseous component follows the same evolution. This
assumption is probably well founded, based on a general agreement
between dust and accretion signatures
\citep[e.g.][]{1995ApJS..101..117K,2006AJ....131.1574L}.

For gas giant formation, the disk lifetime is the strongest
observational constraint. Therefore, any stellar mass dependence on
disk lifetime should be included in planet formation
models. \citet{2007ApJ...660..845B} show this dependence is important,
using it to reproduce a ``valley'' in the period distribution of
planets orbiting stars $>$1.2\,$M_\odot$.

With new discoveries of planets orbiting M-dwarfs and K-(sub)giants,
further trends with stellar mass are emerging. The frequency of giant
planets appears to increase with stellar mass
\citep{2006PASP..118.1685B,2007ApJ...670..833J}. Also, all planets
around stars with masses $>$1.6\,$M_\odot$ have larger ($\sim$1\,AU)
orbits than Solar-mass stars
\citep{2007ApJ...665..785J,2008PASJ...60..539S,2008arXiv0807.0268S}. As
with the valley described by \citet{2007ApJ...660..845B}, the larger
orbits may be caused by changing disk dispersal timescale with stellar
mass.

Recent Spitzer surveys of young clusters have started to reveal
observational evidence of stellar mass dependent disk dispersal: stars
with spectral types earlier than about mid-K appear to lose dust
signatures earlier than their lower mass counterparts
\citep{2006ApJ...651L..49C,2007ApJ...659..599C,2007ApJ...671.1784H,2007AJ....133.2072D}. This
effect has also recently been demonstrated with higher mass Herbig
Ae/Be objects in a number of OB associations
\citep{2005AJ....129..856H} after its initial discovery in 1993
\citep{1993AJ....106.1906H}. Thus, there may be observational evidence
for the stellar mass dependent disk dispersal used in the
\citet{2007ApJ...660..845B} models.

In this paper, we look for evidence of mass dependent disk dispersal
and its evolution using H$\alpha$ equivalent widths (\ha) and
infra-red (IR) excesses in nine $\sim$1--10\,Myr old clusters and
regions. We also compare the disk lifetime inferred from the two
signatures, which probe the different components of circumstellar
disks. We find that disks around higher mass stars tend to disperse
earlier than those around Solar mass stars for all clusters older than
$\sim$3\,Myr. Using all cluster data, we reject the null
hypothesis---that disk dispersal is independent of stellar mass---with
95--99.9\% confidence. The statistical significance of any mass
dependence for individual clusters is not strong, at around the
1\,$\sigma$ level. Using a photoevaporation model, we show why the
signal may be intrinsically weak. We suggest where future observations
can make the most progress towards making our result more
significant. Finally, we return to our motivation and study some
possible effects that stellar mass dependent disk dispersal may have
on giant planet formation. We suggest that mass dependent disk
dispersal may have an observational signature in the semi-major axes
of discovered planets.

\section{Background}\label{sec:background}

Nearly all stars begin their lives with circumstellar disks of gas and
dust \citep[e.g.][]{2000AJ....120.3162L}. The disks have typical radii
of 10-1000\,AU \citep[e.g.][]{1996AJ....111.1977M} and masses
$\sim$0.01--0.1\,$M_\star$
\citep[e.g.][]{2000prpl.conf..559N,2005ApJ...631.1134A}. Assuming the
typical interstellar gas/solids ratio of 100, many disks have material
sufficient to build planetary systems like our Solar System.

Observational probes of disk structure are very sensitive to physical
conditions within the disk \citep[e.g.][]{2007prpl.conf..507N}. The
disk temperature decreases with distance from the central star,
ranging from $\sim 10^3$ to $10^4$\,K close to the star, to 10--50\,K
at larger distances.

At large ($\gtrsim$10-100\,AU) distances, beyond where planets form,
cool gas and dust can be detected through IR and mm
observations. Direct detection of the main gaseous component, H$_2$,
is difficult, so gas disks are inferred from molecular components such
as CO. At all radii, dust is generally inferred from thermal emission,
where detection at longer wavelengths corresponds to cooler dust at
greater distances. For outer disks, the relevant wavelength for dust
emission is in (sub)mm bands.

For inner disks ($\lesssim$10\,AU), where planets probably form,
temperatures and densities are much higher. Direct detection of H$_2$
is still difficult. Gas is usually inferred from evidence of accretion
onto the central star. The most commonly used (and easiest to measure)
accretion signatures are greater than expected UV flux or \ha. These
signatures are caused by shocking/heating of the gas as it accretes
onto the star \citep[e.g.][]{1998ApJ...509..802C,2001ApJ...550..944M}.

The dusty component of the inner disk is detected by an excess of IR
emission above the expected photospheric level at wavelengths beyond
2-3\,$\mu$m. The IR excess is commonly characterised by colour-colour
diagrams, where disks are inferred for objects with redder than
photospheric colours. Alternatively, the shapes of spectral energy
distributions (SEDs) may be used, where disks are inferred for objects
with greater than photospheric SEDs.

Generally, dust and gas signatures agree for young stars; the presence
(absence) of one usually predicts the presence (absence) of the
other. Models suggest that although grains may settle and grow beyond
visible sizes, turbulence and fragmentation ensure the presence of
small grains that remain well mixed with the gas
\citep{2005A&A...434..971D,2008A&A...486..597J}. Therefore, though
disk lifetimes are commonly inferred from dust signatures, there is
observational and theoretical evidence that this method is valid.

There are, however, examples of objects where gas and dust signatures
disagree. Many main-sequence stars show weak mid and far-IR excesses,
but no sign of circumstellar gas. These ``debris disks'' are a
separate class of objects, thought to arise from collisions in
remaining planetesimal belts
\citep[e.g.][]{2004AJ....127..513K}. Around 10\% of stars in young
clusters have little or no excess in the near-IR, but retain large
excesses in mid-IR bands
\citep[e.g.][]{2006AJ....132.2135S,2006AJ....131.1574L}. These
objects, which also have weak or non-existent accretion signatures,
are though to be currently in ``transition'' between the primordial
and debris disk states, with an inner hole as the disk starts to
clear.

Early studies of disk populations in young clusters found a disk
dispersal timescale of $\sim$4--5\,Myr
\citep{2001ApJ...553L.153H,2004ApJ...612..496M}. These studies find
somewhat longer disk lifetimes at longer wavelengths (i.e. greater
distances). This result is unsurprising in the context of grain
growth, because the growth timescale is proportional to the orbital
period. Recent Spitzer surveys have confirmed these timescales with
larger stellar samples.

Increased sample numbers have also allowed the study of how disk
lifetime depends on stellar mass. Studies generally find that the
fraction of stars with disks at a given age changes with the mass of
the host star
\citep{2005AJ....129..856H,2006ApJ...651L..49C,2006AJ....131.1574L,2007ApJ...671.1784H,2007AJ....133.2072D,2008ApJ...675.1375L}. While
nearly all studies find evidence that higher mass stars lose their
disks earlier than Solar mass stars, these results are not
statistically compelling. In addition, \citet{2006ApJ...651L..49C} and
\citet{2008arXiv0801.1116C} find that of objects with IR excesses,
earlier-type stars appear to be in a more evolved state, with smaller
IR excesses relative to the stellar photosphere.

Differences in disk evolution with stellar mass are theoretically
expected. For different stars, observations at fixed wavelength probe
different parts of a circumstellar disk. Because the disk temperature
decreases with distance from the star, thermal emission at longer
wavelengths probes greater distances. For higher stellar luminosities,
regions of fixed temperature (and wavelength) are at greater
distances. Evolutionary timescales such as grain growth depend on
properties such as orbital frequency and gas density, which decrease
with radial distance from the star. Thus, the same wavelength will not
necessarily find a disk in the same evolutionary state for different
stars.

Aside from regions probed by wavebands, more obvious processes exist
that may change the rate of disk evolution with stellar mass. Disks
may experience accelerated photoevaporation due to massive stars in
the local environment
\citep[e.g.][]{2001MNRAS.325..449S,2004ApJ...611..360A,2006ApJ...650L..83B}
or the increased high energy flux of higher mass host stars
\citep{2007ApJ...660..845B}. The effects of the local environment are
most important for low-mass stars. \citet{2004ApJ...611..360A} find
that disks around these stars may be evaporated down to the AU scales
where planets form, particularly in OB associations. The conditions
for external photoevaporation are different for every star in every
cluster and thus may make the presence of global trends for low-mass
stars less likely. In the case of photoevaporation by the host star,
the expected evolution should result in higher mass stars dispersing
their disks earlier, leading to differences in disk fractions for
different mass stars at fixed age. Transition objects can also be
understood in terms of a photoevaporation model, where viscous
evolution and ionising photons from the central star combine to remove
the disk in an inside--out manner \citep[e.g.][]{2001MNRAS.328..485C}.

To study how disk dispersal depends on stellar mass, we compile a
sample of clusters from the literature in \S \ref{sec:data}. To derive
disk fractions, we use IR excesses and \ha. Though these signatures
generally agree, we use both to look for any systematic changes with
cluster age. We first compare the results for overall cluster
fractions, comment on possible effects of stellar multiplicity in \S
\ref{sec:binarity} and study the stellar mass and spectral type
dependence in \S \ref{sec:mass}. Comparing Solar and intermediate-mass
stars, we find evidence for a stellar mass dependence; at about the
2\,$\sigma$ level overall and about 1\,$\sigma$ for individual
clusters.

In \S \ref{sec:models}, we consider different physical mechanisms that
may cause the mass dependence. We first consider grain growth in
different wavebands. Then, using a simple photoevaporation model we
suggest why the mass dependence has low statistical significance. We
argue that an alternative hypothesis, the increasing multiplicity
fraction with stellar mass, is an unlikely cause for mass dependent
disk dispersal.

Finally, we return to the initial motivation for studying disk
evolution and consider some consequences of our results for giant
planet formation in \S \ref{sec:pf}. We suggest that there may be a
signature of stellar mass dependent disk dispersal in the observed
orbits of extra-Solar planets. If the disk dispersal timescale
decreases with increasing stellar mass and the migration timescale is
constant or increases, then above some stellar mass planets will not
have time to migrate before the disk disperses. This effect may cause
the observed outward step in giant planet orbits to $\sim$1\,AU above
1.6\,$M_\odot$.

\section{Cluster data}\label{sec:data}

We select nine well studied clusters and regions from the literature.
To cover a range of disk fractions, our aim is to have clusters with
\ha\ from optical spectra, 3.6--8 $\mu$m photometry from the Spitzer
Infra-Red Array Camera (IRAC), and ages spaced in log time between
1--10\,Myr. Because we want a reasonable number of stars over a wide
stellar mass range, many of our clusters are part of OB associations.

The clusters are: Taurus
\citep{2006ApJ...647.1180L,2006ApJS..165..568F}, Chamaeleon I
\citep{2004ApJ...602..816L,2008ApJ...675.1375L}, IC 348
\citep{2003ApJ...593.1093L,2006AJ....131.1574L}, Tr 37
\citep{2004AJ....128..805S,2005AJ....130..188S,2006ApJ...638..897S,2006ApJ...638..897S},
Upper Scorpius
\citep{1998A&A...333..619P,1999AJ....117.2381P,2002AJ....124..404P,1994AJ....107..692W,2006ApJ...651L..49C},
NGC 2362 \citep{2007AJ....133.2072D}, Orion OB1bc, and OB1a/25Ori
\citep[OB1b and OB1c members combined, and OB1a including 25 Ori
objects,][]{2005AJ....129..907B,2005AJ....129..856H,2007ApJ...661.1119B,2007ApJ...671.1784H},
and NGC 7160
\citep{2004AJ....128..805S,2005AJ....130..188S,2006AJ....132.2135S}. These
clusters populate the 1--10\,Myr age range fairly well when viewed
logarithmically. There is a gap between 2--4\,Myr, which is nicely
occupied by NGC 2264 \citep{2002AJ....123.1528R} and NGC 6611. Though
some data are available \citep{2006ApJ...642..972Y}, the bulk of
Spitzer IRAC photometry for NGC 2264 are unpublished. With a similar
age, the more distant ($\sim$1800\,pc) cluster NGC 6611 shows an
increasing disk fraction with decreasing stellar mass based on near-IR
observations \citep{1993AJ....106.1906H,2005MNRAS.358L..21O}.

These surveys are largely unbiased in terms of disk charateristics,
and/or complete for the range of stars we consider. The lack of bias
mainly results from the requirement that objects have spectral types
to be included in our analysis. The Taurus sample is largely based on
that compiled by \citet{1995ApJS..101..117K}, using IR and
spectroscopic data with no known biases
\citep{1998AJ....115.2491K,2006ApJ...647.1180L}. The current Cha I
sample is thought to be complete to $\sim$0.01\,$M_\odot$
\citep{2007ApJS..173..104L}, well below masses we consider. However,
objects with measured \ha\ may be biased due to the wide range of
techniques used to classify objects known when the \ha\ measurements
were made \citep{2004ApJ...602..816L}. The IC 348 sample is thought to
be 100\% complete to 0.03\,$M_\odot$ \citep{2003ApJ...593.1093L}. The
Orion, Tr 37, and NGC 7160 samples are based on variability and thus
unbiased in terms of disk signatures \citep{2001Sci...291...93B}. The
Upper Sco sample has been compiled to ensure an unbiased sample
\citep{2001AJ....121.1040P}. The NGC 2362 is the only questionable
cluster we use, thought to be complete to $\sim$0.5\,$M_\odot$. The
sample has been compiled using a number of methods, including an
H$\alpha$ survey \citep{2005AJ....130.1805D}, which may bias the
results in favour of stars with disks for the lowest-mass stars. Any
bias is probably minimal, as disk fractions from both accretion and
dust signatures are found to be similar to the other 5\,Myr old
regions, Upper Sco and Orion OB1bc \citep[][this
paper]{2007AJ....133.2072D}.

\begin{figure*}
\plotone{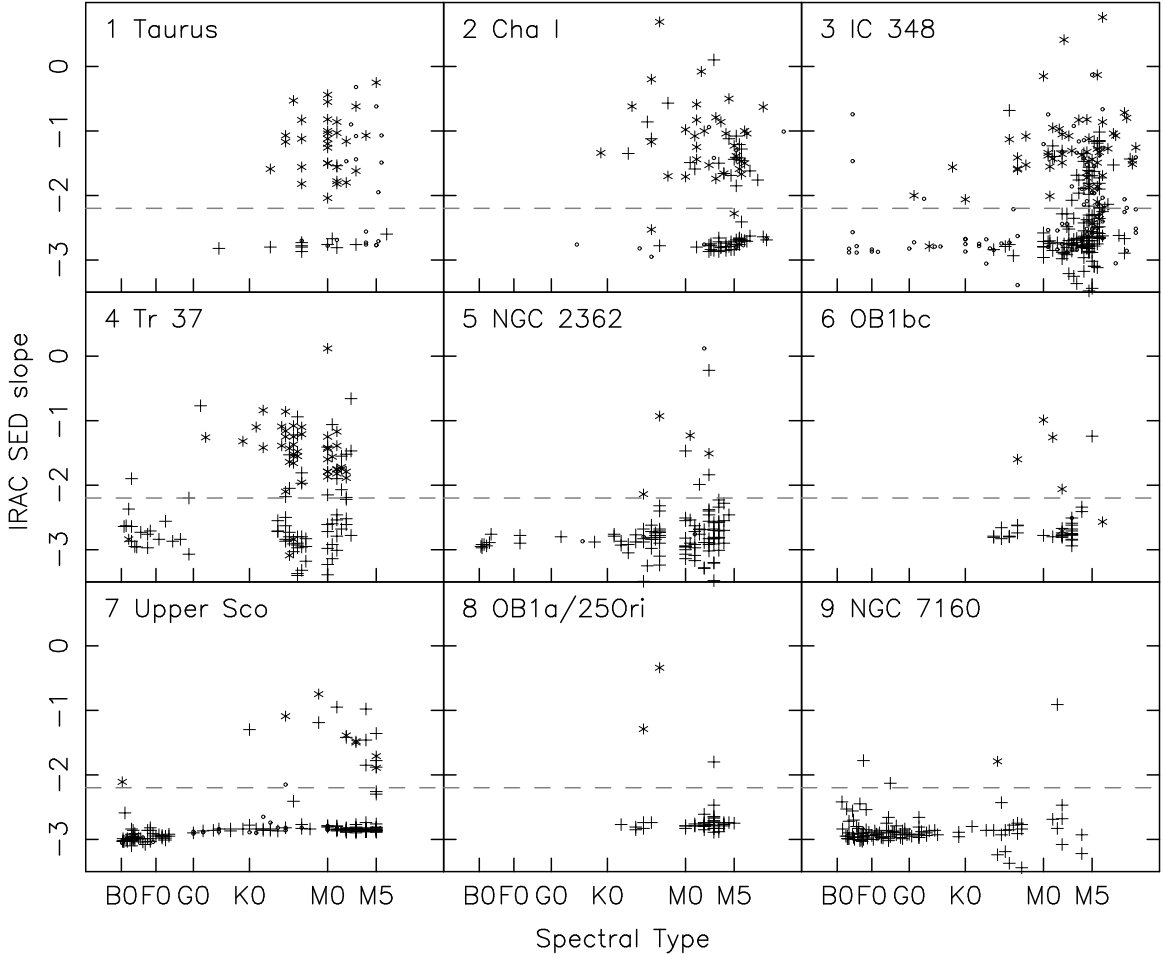}
\caption{IRAC SED slopes for objects with spectral types. Symbols show
  accreting (*), non-accreting (+), and objects without \ha\
  measurements ($\circ$). The x-axis is expanded toward later spectral
  types for clarity. Based on these data, we use a slope of $\alpha >
  -2.2$ (dashed line) to distinguish between stars with and without
  dust disks. Orion OB1bc lacks SED slopes for intermediate mass stars
  because their disk classification is based on JHK-excesses
  \citet{2005AJ....129..856H}.}\label{fig:sed}
\end{figure*}

To distinguish between stars with and without disks, we use two
standard measures: \ha\ and IR excesses. When high resolution data are
available, we use the spectral type independent criterion of 10\%
H$\alpha$ width $> 270$\,km s$^{-1}$ to distinguish accreting (CTTS)
and non-accreting (WTTS) stars. For the more common low resolution
spectra, we use the spectral-type dependent accretion criterion of
\citet{2003ApJ...582.1109W}, which accounts for different
chromospheric levels of H$\alpha$ for different spectral
types. \citet{2006AJ....132.2135S} find good agreement between \ha\
derived from high and low resolution spectra with Tr 37, but note that
low resolution spectra may become less reliable for older clusters as
accretion rates drop. We add a lower threshold of \ha\ $> 0$ for
early-type stars $<$G9 (theirs is \ha\ $> 3\,\dot{A}$ for all stars
$<$K6), an approximate effective temperature where main-sequence
dwarfs start to show H$\alpha$ in absorption
\citep[e.g.][]{2003IAUS..210P.A20C,2005AJ....129..856H}. The few
early-type accreting objects tend to have relatively large \ha\, so
our results are not sensitive to the criterion for $<$G9 stars.  We
refer to objects with excess \ha\ as accretors. The disk fraction thus
derived is the ``accretion fraction.''

To provide a measure of the IR excess from a dust disk, we use the
slope of the SED, defined by $\alpha = {\rm d} \log \lambda F_\lambda
/ {\rm d} \log F_\lambda$
\citep[e.g.][]{1987ApJ...312..788A,2006AJ....131.1574L}. Most young
stars have relatively shallow or flat SEDs ($\alpha \gtrsim -2$) that
are easily distinguished from the steep SEDs of the typical stellar
photosphere ($\alpha \lesssim -2.5$). Thus, the slope of the SED is a
simple way to distinguish stars from disks. To take advantage of the
large spatial coverage of most Spitzer surveys, we use IRAC data to
derive $\alpha$ for stars with data in all four bands ([3.6], [4.5],
[5.8], and [8]). Magnitudes are dereddened using the relation derived
by \citet{2005ApJ...619..931I}. For Upper Sco, we use the slope
between [4.5] and [8] from \citet{2006ApJ...651L..49C}, which
generally agrees with their 8\,$\mu$m disk classification. For
intermediate-mass HIP objects in Orion OB1bc, we use the JHK disk
classification from \citet{2005AJ....129..856H}.

\begin{figure*}
\plottwo{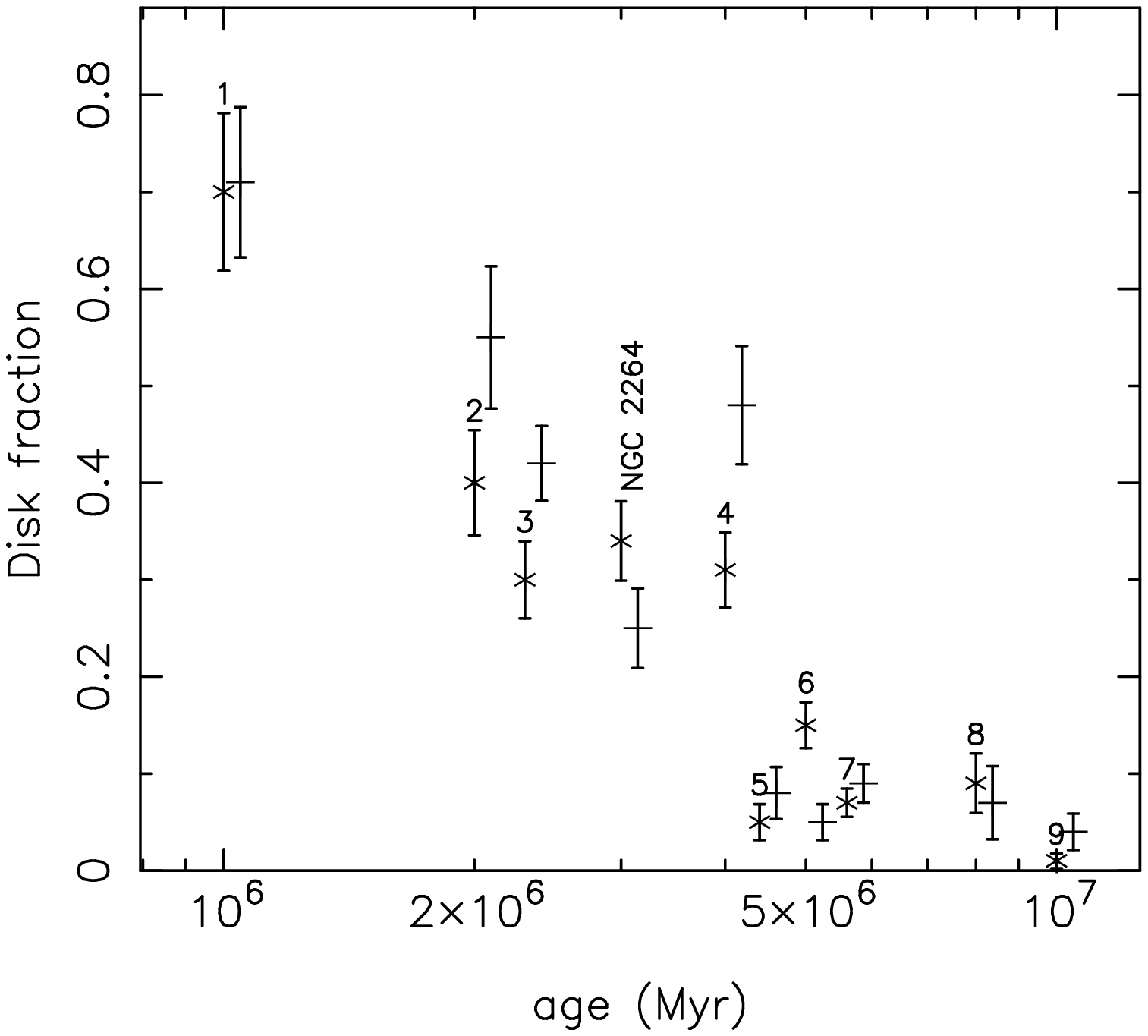}{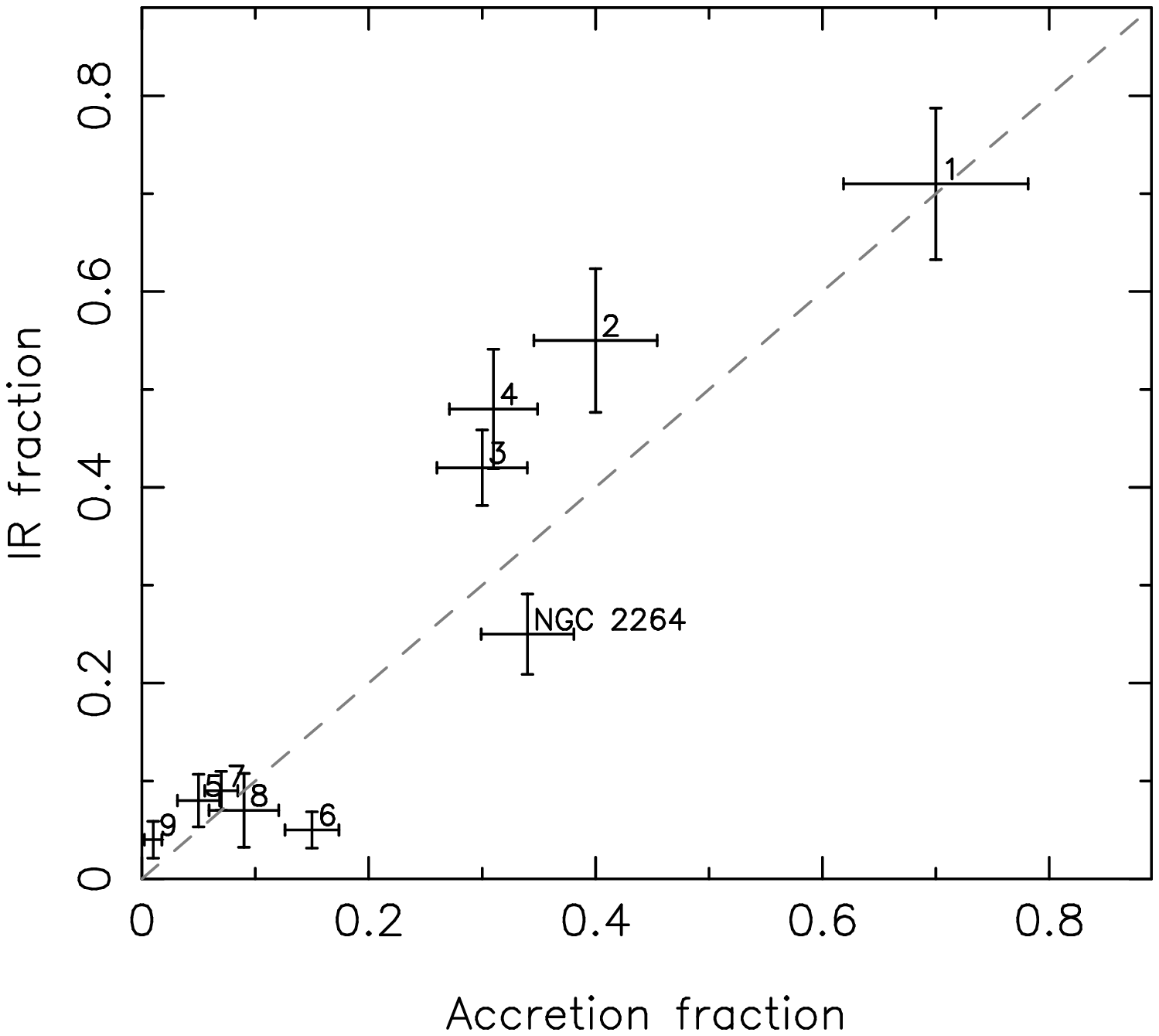}
\caption{Overall cluster accretion ($\times$) and IR-fractions (+), as
  a function of cluster age (left), and compared (right). For clarity,
  IR fractions are offset slightly right of adopted cluster ages,
  Upper Sco is offset right and NGC 2362 to the left of 5\,Myr. In the
  right panel, equal fractions lie on the dotted line. Errors are
  Poisson ($\sqrt{N}$) estimates and the systematic error in age is a
  few Myr \citep[e.g.][]{2001ApJ...553L.153H}.}\label{fig:tot-fr}
\end{figure*}

Figure \ref{fig:sed} shows dereddened IRAC SED slopes for our cluster
sample. There is a clear divide between objects with and without disks
for the youngest clusters, Taurus and Cha I. The separation between
objects with and without disks is least clear for IC 348: the IR
fraction varies between 30\% for $\alpha = -1.8$, to 57\% for $\alpha
= -2.6$. This variation results from a decreasing median SED slope
with cluster age and/or the presence of transition objects (see
below), which make the divide less clear for older clusters
\citep[e.g.][]{2007ApJ...671.1784H}. For a brief discussion of the
limitations of using the SED slope, see \citet{2007AJ....133.2072D}.

Based on Figure \ref{fig:sed}, we choose $\alpha > -2.2$ to
distinguish between stars with and without disks. Our results do not
vary significantly for reasonable range ($-2.6 < \alpha < -1.8$). We
do not distinguish between stars with ``weak'' or ``anemic'' disks
\citep[e.g.][]{2006AJ....131.1574L} and those with photospheric SED
slopes. The disk fractions thus derived are termed ``IR fractions.''

Requiring a star to have both accretion and dust indicators ensures a
robust classification of objects with primordial disks. However, this
constraint precludes any check on whether the dusty and gaseous
components evolve together and reduces our sample numbers considerably.
Thus, we consider accretion and IR indicators individually and compare 
the results to those for disks with both signatures.

Before analysing the cluster data in detail, we first consider overall
disk fractions for each cluster. To provide data at 3\,Myr, we include
results from NGC 2264 for this figure only. The IR fraction for this
cluster is derived from the I-K colour
\citep{2002AJ....123.1528R}. Though the I-K disk indicator is at
shorter wavelengths, it gives a rough estimate of the disk fraction
that we expect from Spitzer IRAC data.

Figure \ref{fig:tot-fr} shows disk fractions derived from the overall
cluster data. Disk fractions decay from $\sim$70\% at 1\,Myr to
5--10\% by $\sim$5--10\,Myr. Tr 37 has a relatively high IR fraction
($\sim$50\%) for its age; however, this fraction is not unreasonably
high given the scatter observed for other clusters. Though the
systematic uncertainty in ages is a few Myr, ranking the clusters by
their disk fractions yields nearly the same order as ranking the
clusters by their ages. Clusters lose most of their disks in
$\sim$5\,Myr, with a small fraction or stars retaining disks to
$\sim$10\,Myr. Disk fractions from IR excesses and \ha\ generally
agree (right panel of Fig. \ref{fig:tot-fr}), showing that decay
timescales for the gaseous and dusty components are similar.

\subsection{Binary and Multiple Systems}\label{sec:binarity}

The disks of stars in binary systems evolve differently from
those around single stars. A companion star truncates the
circumstellar disk, shortening the disk lifetime. The extent to which
the disk is truncated depends on the separation and mass ratio
\citep[e.g.][]{1977MNRAS.181..441P,1999MNRAS.304..425A}. Circumbinary
disks may also be present. Therefore, the extent to which a binary
companion dominates disk evolution is set by the parameters of each
individual system.

Given the natural range of mass ratios and separations, disk dispersal
in binary systems is complex. As a further complication, the
occurrence of binaries is a function of the mass of the primary
\citep[e.g.][]{2006ApJ...640L..63L}. Thus, for some fixed separation
distribution, disks may disperse earlier on average for higher mass
stars, depending on the prevalence of circumbinary disks for the
closest systems. If enough systems have close orbits and the change
in binary fraction with stellar mass is strong enough, this process
may cause an observable difference in disk fractions over a range of
stellar masses.

% SELECT * FROM stars WHERE cluster = 'Upper Sco' AND id_HIP IS NOT
% NULL AND bin_binary = 'yes'

To look at the effects of companions, we construct two subsamples of
objects with ``known'' multiplicity. The first consists of the
\citet{2006ApJS..165..568F} Taurus sample, for which we compare
multiple systems with the balance of objects. Multiplicity in Upper
Sco has also been well studied
\citep[e.g.][]{1987ApJS...64..487L,2002A&A...382...92S,2005A&A...430..137K,2007A&A...464..581K}.
We collect binary and multiple objects observed and compiled by
\citet{2007A&A...474...77K}, objects flagged as X, O, or G in the
Hipparcos catalogue \citep[see][]{2007A&A...474...77K}, and objects
from the Catalog of Components of Double \& Multiple stars (CCDM)
\citep{2002yCat.1274....0D}.\footnote{Vizier catalogs: Hipparcos
  I/239/hip\_main; CCDM I/274/ccdm} We then compare the disk fraction
of systems to all other Upper Sco objects with Hipparcos
identifiers. Due to the aforementioned bias, the real multiplicity
fraction in both associations is almost certainly larger than
observed. For higher mass stars, such as the BA-type Hipparcos stars
in Upper Sco, the multiplicity fraction may be near unity
\citep{2007A&A...474...77K}. There are too few objects to make any
useful comparison of the effect of different companion separations.

\begin{figure*}
\plotone{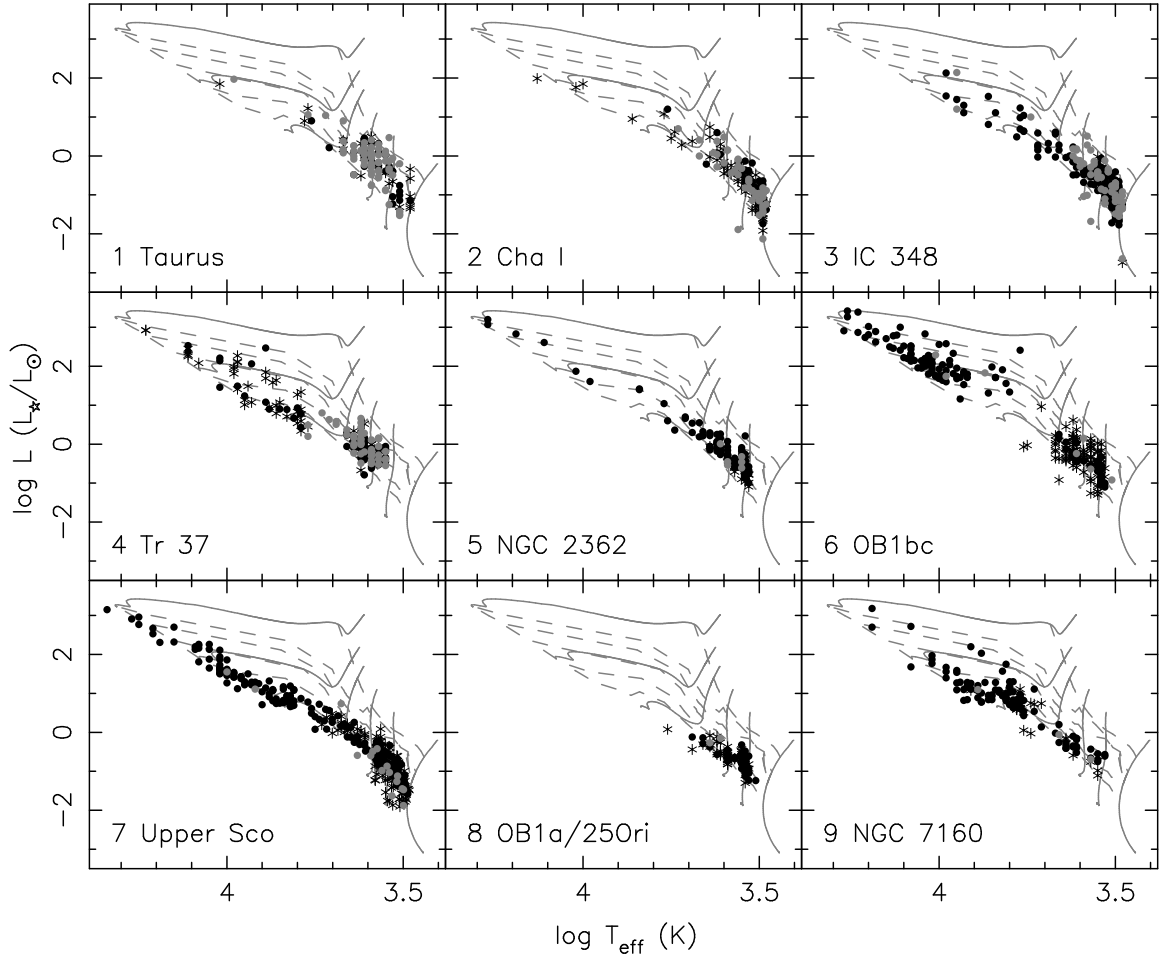}
\caption{HR diagrams of objects with (grey filled circles), and
  without (black filled circles) IR excesses in our clusters. Objects
  without IR classification (which may be CTTS/WTTS) are marked by
  *'s. Grey lines show \citet{2000A&A...358..593S} PMS tracks for 0.1,
  0.3, 0.6, 1.5, 3, and 7\,$M_\odot$ stars (solid lines) and
  isochrones for 0.1, 0.5, 1, 2, 5, and 10\,Myr (dashed
  lines).}\label{fig:hr}
\end{figure*}

Table \ref{tab:binary} shows the disk fractions of single and binary
objects for our two samples. For both Taurus and Upper Sco there is no
apparent difference in disk fractions between single and multiple
stars. In other samples, binaries with separations $\gtrsim$20\,AU
appear to have little impact on disk lifetime and evolution
\citep[e.g.][]{2007prpl.conf..395M,2008ApJ...673..477P}.  Thus, we
conclude that binaries do not impact our derived disk fractions.

% SELECT * FROM stars WHERE cluster = 'Upper Sco' And id_HIP is not null
% and bin_2_per_d is not null

Exploring the possible impact of multiplicity on disk evolution merits
further study. Samples with a large range in separation and primary
mass are needed to study the possible implications for stellar mass
dependent disk dispersal. However, based on the results of Table
\ref{tab:binary} and other observational studies
\citep{2006ApJS..165..568F,2007prpl.conf..395M,2008ApJ...673..477P},
we do not exclude known multiple systems from our sample.

\subsection{Stellar mass dependence}\label{sec:mass}

We now use our clusters to look for stellar mass dependent disk
dispersal. In each cluster we split stars into bins defined by
spectral type and stellar mass and examine the resulting disk
fractions. We first qualitatively study the data and find that some
stellar mass dependence appears for Solar and intermediate-mass
stars. We then focus on these stars and quantify the significance of
the dependence in several independent ways.

To obtain the additional information required to analyse each object
in our database, we obtain spectral types and extra photometry from
Simbad\footnote{http://simbad.u-strasbg.fr/simbad/}
%\footnote{\href{http://simbad.u-strasbg.fr/simbad/}{http://simbad.u-strasbg.fr/simbad/}}
and Vizier.\footnote{http://webviz.u-strasbg.fr/viz-bin/VizieR}
%\footnote{\href{http://webviz.u-strasbg.fr/viz-bin/VizieR}{http://webviz.u-strasbg.fr/viz-bin/VizieR}}
Where needed, we calculate extinction using the dwarf colours of
\citet{1995ApJS..101..117K} and conversions from
\citet{1988PASP..100.1134B} and \citet{1989ApJ...345..245C}. We
exclude objects with no spectral type. For all possible stars, we
derive effective temperatures from the spectral type
\citep{1995ApJS..101..117K} and luminosity using dereddened J or I
magnitudes.

\begin{figure*}
\plottwo{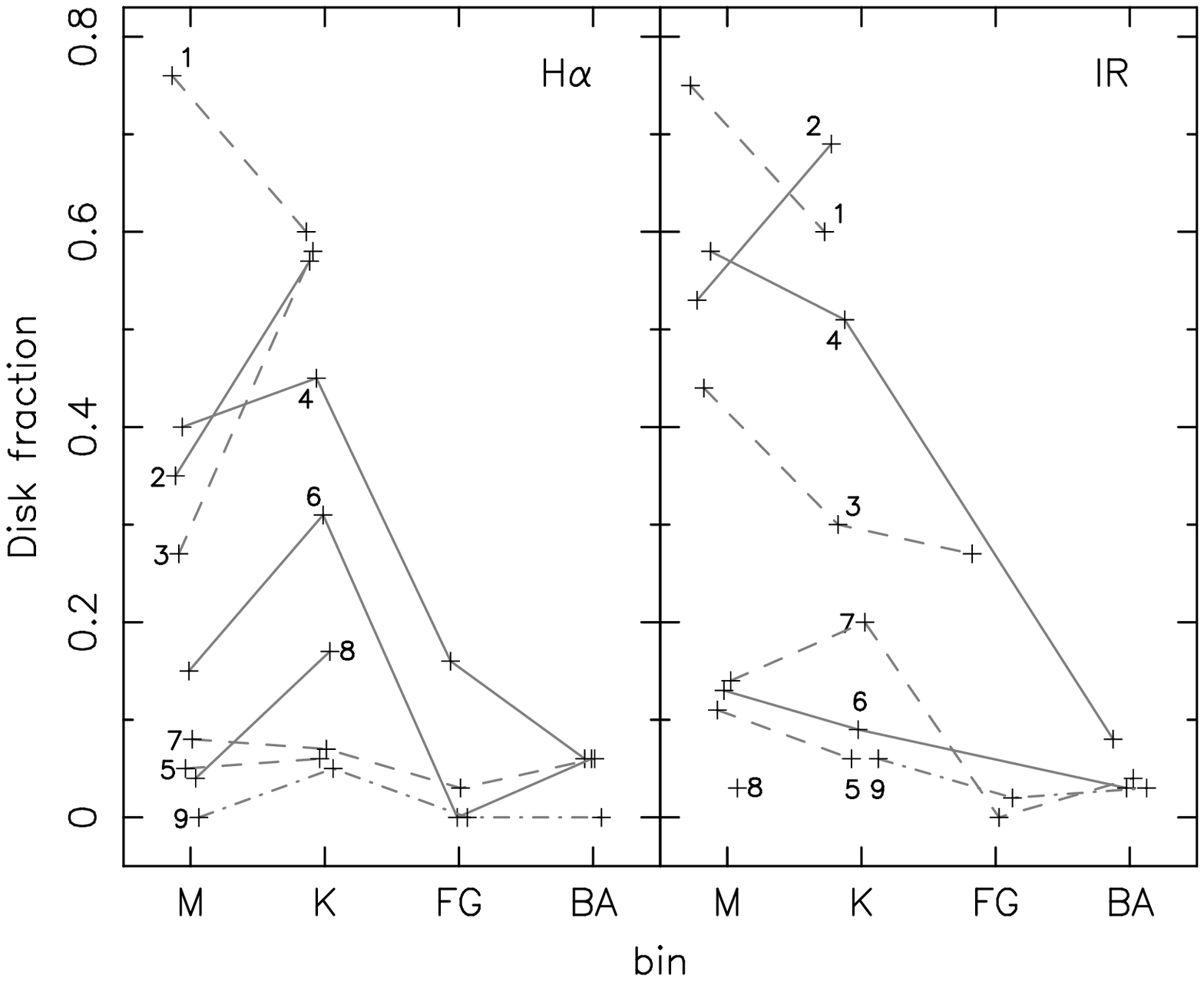}{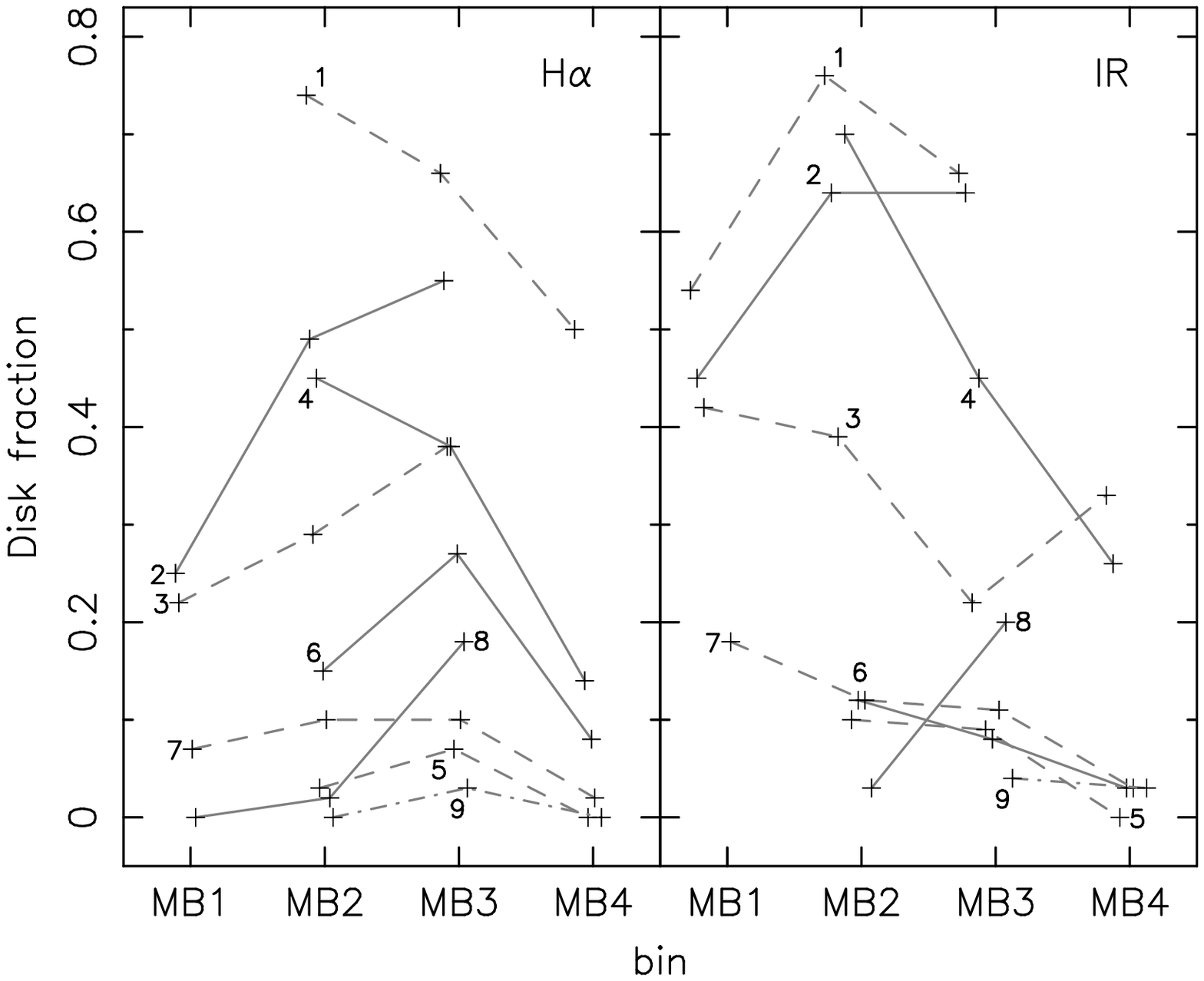}
\caption{Accretion and IR disk fractions binned by spectral type
  (left) and mass (right). Grey solid, dashed, and dot-dashed lines
  (the same in each panel and Figure \ref{fig:bin-both-fr}) link
  points from each cluster, which are numerically labeled in the same
  order as Table \ref{tab:data-spty}. Bins with less than 10 stars are
  omitted and for clarity, points are offset left and right from their
  bin locations slightly. Errors are omitted for clarity, but are
  included in Figure \ref{fig:bin-ir-compare}.}\label{fig:bin-fr}
\end{figure*}

We begin by showing HR diagrams for the stars in our clusters in
Figure \ref{fig:hr}. For most clusters, stars appear reasonably evenly
distributed in mass. Orion OB1bc is a notable exception: the CIDA
variability study of Orion \citep{2005AJ....129..907B} was limited to
lower mass stars, apparently due to CCD saturation for brighter
objects. There is a marked gap between these objects and the higher
mass Hipparcos objects studied by \citet{2005AJ....129..856H}. Many of
the higher mass stars in the \citet{2007AJ....133.2072D} NGC 2362
sample lack spectral types. Some deficiency of $\sim$G-type objects is
also expected due to stellar evolution, when $\sim$Solar-mass stars
develop a radiative core and move to the main-sequence at roughly
constant luminosity.

Figure \ref{fig:hr} also identifies stars with IR excesses (grey
dots). The decline in overall disk fraction with cluster age can be
seen in the change from mostly grey to nearly all black (no disk) data
points. For the youngest clusters, stars with and without disks appear
evenly distributed.  For older clusters, however, stars with disks
have a marked spatial dependence on where stars lose their
disk. Although many low mass stars in Tr 37 have disks, there are no
stars with $\log T_{\rm eff} > 3.8$ (or $\log L_\star / L_\odot > 1$)
with disks. Other clusters have a similar, but less obvious, lack of
disk signatures among more massive stars.

To quantify how the disk fraction changes with stellar mass or
spectral type, we bin the data. Some previous studies bin objects by
their spectral type
\citep[e.g.][]{2006ApJ...651L..49C,2007ApJ...671.1784H}. However, for
the wide range of stellar masses we consider, pre--main-sequence (PMS)
tracks have different loci in the HR diagram. Convective low mass
stars follow vertical Hayashi tracks at constant $T_{\rm eff}$.  As
they develop a radiative core, intermediate mass stars increase in
$T_{\rm eff}$ at nearly constant luminosity. Thus, binning by spectral
type is roughly a mass bin for low mass stars and an age bin for
intermediate mass stars. The opposite is true for binning by
luminosity. If a cluster contains stars with an apparent range of
ages, earlier spectral types contain systematically older stars for a
given stellar mass. This bias may result in an artificially high
difference in disk fractions between bins.

\subsubsection{Binned Data: Qualitative results}

We bin our clusters by spectral type (M, K, FG, and BA) and by mass
(0.1--0.3, 0.3--0.6, 0.6--1.5, and 1.5--7\,$M_\odot$) using the
\citet{2000A&A...358..593S} PMS tracks. We name the mass bins MB1,
MB2, MB3, and MB4 respectively. Objects lying slightly below the main
sequence are included in the nearest mass bins by eye (e.g. Upper Sco
intermediate mass stars). Bins MB1 and MB2 correspond well to M-type
stars of all ages. As noted above, stellar evolution means that the
correspondence for MB3 and MB4 changes with age. For young clusters,
MB3 corresponds to K-type stars and MB4 to BAFG-types. By 10\,Myr,
MB3 contains GK-types and MB4 BAF-types.

The results of binning stars by mass and spectral type are summarised
in Tables \ref{tab:data-spty} and \ref{tab:data-mass}. Also shown are
assumed ages and distances. Percentages next to each mass bin label in
Table \ref{tab:data-mass} show the fraction of stars expected for a
standard IMF \citep{2001MNRAS.322..231K}. The lowest mass and spectral
type bins are incompletely covered for some clusters
(Fig. \ref{fig:hr}), though this should not affect disk
fractions. Though different PMS models differ significantly, our
results do not differ much with other tracks
\citep[e.g.][]{1999ApJ...525..772P}, because they have the same
general form (i.e. stellar mass increases with $T_{\rm eff}$ and/or
$L_\star$ and we are free to choose any mass binning).

Figure \ref{fig:bin-fr} shows accretion and IR fractions for our
chosen bins. The binned data show a systematic decrease in disk
fraction with increasing stellar mass and spectral type. Both
accretion and IR fractions decrease from the K/MB3 to FG-BA/MB4 bins,
for all clusters older than $\sim$2\,Myr old IC 348.

If the decrease in disk fraction with increasing stellar mass were
only present in \ha\ measurements, it might be attributed to a
detection bias against early-type stars, where excess H$\alpha$ is
harder to detect in low resolution spectra. However, it is present in
both the accretion and IR fractions, and based on the general
agreement between the two signatures is probably real.

\begin{figure*}
  \plottwo{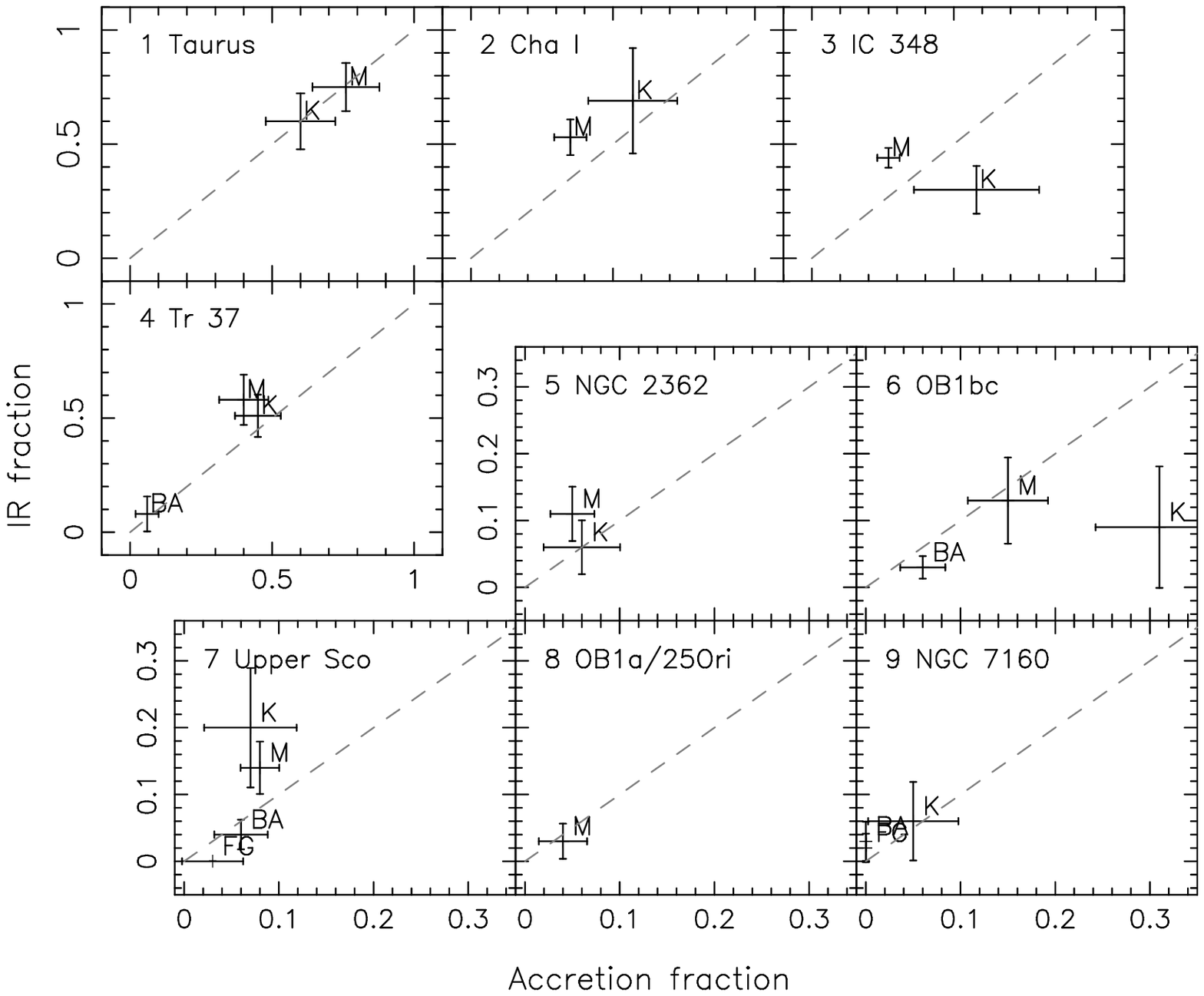}{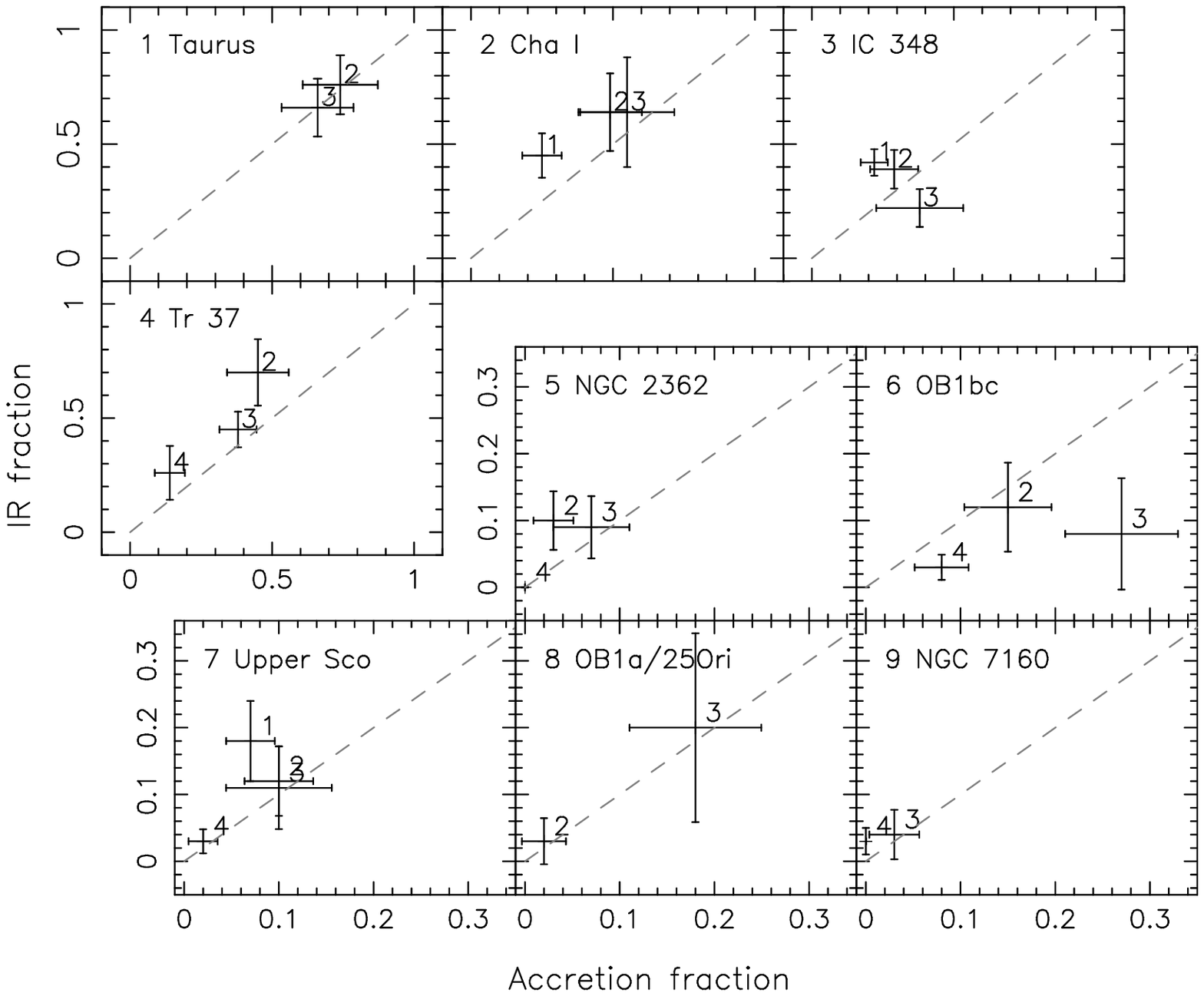}
  \caption{Accretion and IR disk fractions compared as in
    Fig. \ref{fig:tot-fr}, but binned by spectral type (left) and mass
    (right). Errors are Poisson estimates and labels refer to the bin
    ($1 = {\rm MB1}$ etc. for mass binned data). The lower right set
    of subpanels have zoomed axes for lower disk fractions in older
    clusters.}\label{fig:bin-ir-compare}
\end{figure*}

We observe, and expect, less stellar mass dependence on disk fraction
for the youngest clusters. If stars of all spectral types have disks
at early stages, then a difference in disk fractions between bins has
had little time to develop for the youngest clusters. The results for
these clusters will be somewhat influenced by the small numbers of
intermediate-mass stars. The disk fraction in Cha I increases with
increasing stellar mass \citep{2008ApJ...675.1375L}, which may be the
result of initial variations in disk fraction with stellar mass. Only
a few intermediate mass stars in the IC 348 sample of
\citet{2003ApJ...593.1093L} have \ha\ measurements, so the high
accretion fraction for these objects may simply be due to the small
sample size. It is not clear why the IR fraction of IC 348 increases
from MB3 to MB4. Given that the disk fraction is lower than for Tr 37,
we expect it to show a similar trend. The difference may be a result
of its younger age and perhaps an initially lower disk fraction.

Compared to Solar and intermediate-mass stars, any trends for low-mass
and late spectral type stars are less clear. There appears to be a
consistent decrease in accretion fraction to the lowest mass/spectral
type bins, but this trend does not appear in the IR-fractions as we
bin them. Finer binning shows some clusters do decrease their
IR-fractions toward low mass stars, though the uncertainties are large
\citep[e.g.][]{2007AJ....133.2072D,2007ApJ...671.1784H}. 

External photoevaporation may affect low mass stars
\citep[e.g.][]{2001MNRAS.325..449S,2004ApJ...611..360A}, particularly
because many of our clusters are part of OB associations. Given that
external photoevaporation depends on the cluster environment and is
therefore different for each star in each cluster it may contribute to
the lack of a clear trend.

Looking more closely at the objects with disagreement between their IR
and accretion signatures yields information about objects that are
likely transition disks. This comparison applies to individual objects
with both IR and \ha\ measurements. Of the objects with disagreeing
signatures, 43 are accretors without IR excesses, and 109 have IR
excesses but are not accretors (i.e. may be transition disks). The
former group are spread over spectral type bins roughly in proportion
to the overall sample, and may be due to near-threshold/erroneous
values of \ha\ or SED slope. However the latter group, the transition
disks, are nearly all K and M-type stars. About 80\% (87/109) of these
transition objects are M stars (vs. 50\% (650/1253) of the overall
sample), and 16\% (18/109) are K stars. Thus, the apparent discrepancy
between disk fractions for lower-mass stars may be partly accounted
for by nearly all transition disks being lower-mass stars.\footnote{A
  similar finding for stars with ``anemic'' disks---those with
  near-photospheric SED slopes---in IC 348 was made by
  \citet{2006AJ....131.1574L}, and interpreted as being due to faster
  disk evolution for later-type stars. However,
  \citet{2008arXiv0801.1116C} finds the converse for this sample,
  suggesting that anemic disks around early-type stars are in fact
  debris disks and further evolved than the transition disks around
  the late-type stars.}

Given the lack of any clear trend when comparing low and Solar-mass
stars, we now focus on trends for Solar and intermediate-mass stars.

Figure \ref{fig:bin-ir-compare} again shows the binned results, this
time comparing accretion and IR fractions. The error bars provide a
measure of the significance of the differences seen in Figure
\ref{fig:bin-fr} and show no major differences between IR and
accretion fractions. The three young clusters Taurus, Cha I, and IC
348 have disk fractions that are largely consistent with one
another. Of the older clusters, Tr 37, Orion OB1bc, and Upper Sco have
somewhat significant differences between the K/MB3 and BA-FG/MB4
bins. Though the individual significance is not consistently high, of
the older clusters with intermediate-mass stars, all five have lower
disk fractions in their highest mass and spectral type bins.

\begin{figure}[b]
\plotone{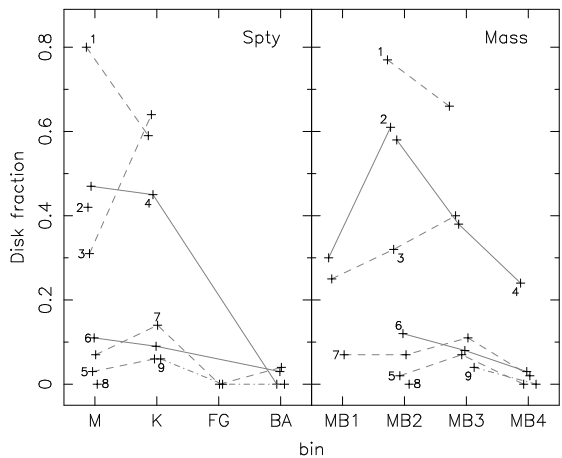}
\caption{Same as the left panels of Figure \ref{fig:bin-fr}, but for
  stars showing both accretion and dust signatures. Errors are omitted
  for clarity, but are larger than in Figure \ref{fig:bin-ir-compare}
  due to fewer objects.}\label{fig:bin-both-fr}
\end{figure}

To try a more robust primordial disk classification, we require a star
to have both accretion and IR signatures to be classed as having a
disk, because objects with only one disk signature may be some form of
transition object. Figure \ref{fig:bin-both-fr} shows these ``common''
disk fractions. There are fewer objects overall, but the primary
result is the same: the decrease in disk fraction for higher
mass/later spectral type bins remains.

\begin{figure*}
\plottwo{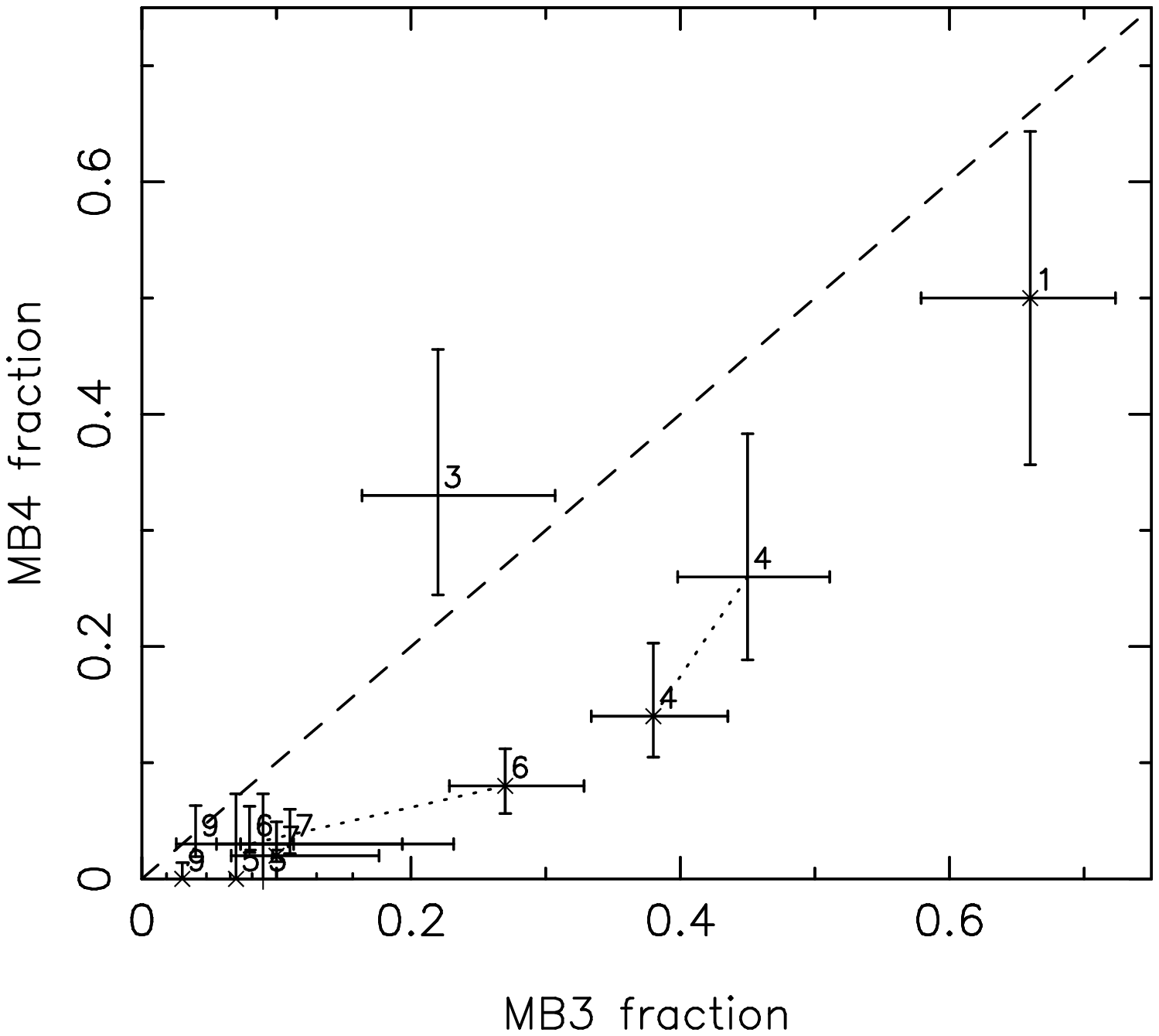}{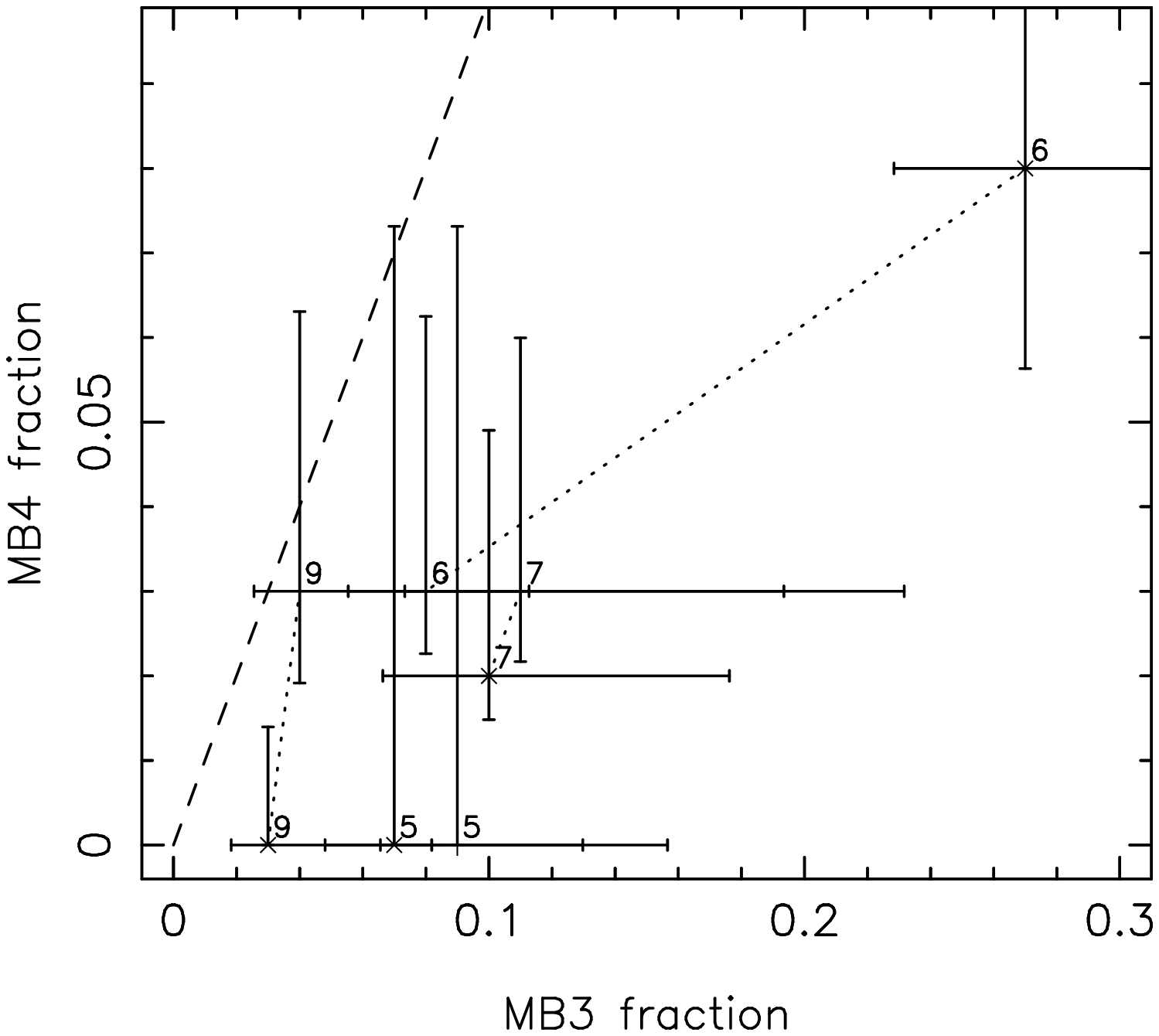}
\caption{Accretion ($\times$) and IR (+) disk fractions for stellar
  mass bins MB3 and MB4. Note the different scales in each plot: the
  right panel shows detail near the origin for $>$5\,Myr
  clusters. Bins with equal fractions lie near the dashed line. The
  errors are 1\,$\sigma$, as described in the text. Where a bin has no
  disks, the error with 1 disk is used for the upper
  error.}\label{fig:irac}
\end{figure*}

There are some differences between the results for binning by spectral
type and mass. For mass binning, general trends are conserved, but the
magnitude of the differences is similar or smaller. The difference in
disk fraction between K and BA-stars in Tr 37 is $\sim$40\%, but only
$\sim$20\% for the roughly comparable MB3 and MB4
stars. \citet{2006ApJ...651L..49C} find that the disk fraction of KM
stars in Upper Sco is higher than FG and BA stars at 99.2\% and 92\%
confidence respectively. For MB3 and MB4 stars, we find a similar
result of 87--90\% confidence (see Table \ref{tab:fet} and discussion
below). Because it has a physical basis, we proceed using the mass
binned results.

\subsubsection{Quantifying results for Solar and intermediate-mass
stars}\label{sec:quant}

We now put the differences in disk fractions on a more quantitative
footing. There are two aspects of the results we can quantify: the
significance of differences between bins on a cluster by cluster
basis and the significance yielded by our sample of nine clusters.

Figure \ref{fig:irac} shows a direct comparison of the MB3 and MB4 bin
fractions (where both bins have $>$10 stars). The errors are
calculated from a binomial distribution and represent 1\,$\sigma$
limits on the intrinsic disk fraction, given the observed fraction
\citep{2003ApJ...586..512B}. The most obvious result is that all
clusters, with the exception of IC 348, have lower disk fractions in
the higher mass bin (Cha I is also an exception, but has only 3 (6)
MB4 stars with IR (accretion) measures. OB1a also has insufficient MB4
stars to feature on this plot). These data may suggest that only older
clusters tend to show lower disk fractions in MB4 than MB3. Perhaps
primordial variations in disk fraction with stellar mass are too large
for clusters to show consistent results until most of the disks have
been dispersed.

To test the significance of the trend for our clusters, we compute the
$\chi^2$ value for the null hypothesis that MB3 and MB4 have equal
disk fractions---that disk dispersal is independent of stellar
mass. Using the 12 data points (Fig.~\ref{fig:irac}) with $\ge$10
stars and errors added in quadrature, we find $\chi^2 = 19.5$ for
perpendicular deviations. We therefore reject the null hypothesis at
95\% confidence. This relatively low confidence reflects the errors
associated with each cluster.

Though the measurements of accretion and IR signatures are
independent, they are correlated because stars with one signature tend
to show the other. Thus there are fewer than the 11 degrees of freedom
used above. If we combine the accretion and IR fractions of the
clusters, by taking the average number of disks and stars for the two
signatures, we find $\chi^2 = 23$ for 7 data points (Cha I and OB1a
still have less than 10 stars in MB4). We reject the null hypothesis
with 99.9\% confidence. Though the differences to the null hypothesis
are similar, the rejection is much stronger because the errors for
each point are smaller. Therefore we conclude that there is some
evidence of stellar mass dependent disk dispersal with 95--99.9\%
confidence. We return to the question of whether the data can be
better explained by a model in \S \ref{sec:pe}.

On an individual level, these plots make the differences between disk
fractions clear. Within their errors, Taurus, IC 348, and NGC 7160 are
all consistent with having equal disk fractions in MB3 and MB4. We
expect this result, because at early and late times, all or no stars
have disks. Tr 37 appears to have the most significant deviations from
equal disk fractions, with both accretion and IR fractions well
below the dashed line in Figure \ref{fig:irac}. Looking at clusters
with low overall disk fractions (right panel), NGC 2362, Orion OB1bc,
and Upper Sco also have somewhat significant deviations from equal
fractions.

We test the significance of individual differences between MB3 and MB4
with Fisher's Exact Test (right-sided). This test finds the likelihood
that a more extreme separation in disk fractions between the two bins
should exist. In the right sided case, ``more extreme separations''
means a more positive difference in disk fractions ${\rm MB3} - {\rm
  MB4}$. This test shows the significance of the difference in each
cluster and highlights where future observations can make the most
progress.

Table \ref{tab:fet} shows the results of this test for accretion, IR,
and common fractions, for objects in MB3 and MB4. The numbers in
parentheses show the lower of the total number of stars in MB3 or MB4
for each cluster, as a measure of the minimum sample size each test
result is based on. Thus, 0.19\%(50) for Tr 37 is derived from
comparing 34/89 (MB3) with 7/50 (MB4).

The accretion fractions of Tr 37 and OB1bc are by far the most
significant, with less than 0.2\% likelihood that a greater difference
in disk fractions between MB3 and MB4 should occur. These clusters
have the most significant results because of the large number of
intermediate-mass stars with \ha\ measurements. By eye, this test
agrees with the separations suggested by the errors in Figure
\ref{fig:irac}. As noted above, we find a similar significance to
\citet{2006ApJ...651L..49C} for Upper Sco. Thus, this test shows that
the significance of differences in MB3 and MB4 disk fractions for the
4--5\,Myr old clusters vary between 60--99\%. It is this large range
in significance that limits the confidence with which we reject the
null hypothesis.

As can be intuitively seen in Figure \ref{fig:irac}, Fisher's Exact
Test shows where the most progress can be made by future
observations. The Spitzer observations of Orion OB1a/25 Ori and OB1bc
\citep{2007ApJ...671.1784H} focus on low-mass stars
\citet{2005AJ....129..907B}. Many intermediate-mass stars have yet to
be characterised, particularly in the Spitzer IRAC wavelength range we
use here. NGC 2362 has a large population of intermediate-mass stars
lacking published spectral types, which would increase the
significance of the result for this cluster.

\subsection{Summary}

We find evidence of stellar mass dependent disk dispersal. All
intermediate-mass stars in clusters older than $\sim$3\,Myr have lower
disk fractions than Solar-mass stars. Individually, the 4--5\,Myr
clusters Tr 37, NGC 2362, Orion OB1bc, and Upper Sco all have marginal
significance, at around the 1\,$\sigma$ level. The most significant
results are for the accretion fraction of Tr 37 and Orion OB1bc. These
results are generally in agreement with the results of the Fisher's
Exact Test. We reject the null hypothesis---that the disk fractions in
bins MB3 and MB4 are the same---at 95--99.9\% confidence. This
confidence is limited by the number of stars in each cluster.

\section{Theoretical mechanisms}\label{sec:models}

The results in Figure \ref{fig:irac} and Table \ref{tab:fet} suggest
that mass dependent disk dispersal is real. Though intermediate-mass
stars in all clusters older than $\sim$3\,Myr have lower disk
fractions than Solar-mass stars, the results of Fisher's Exact Test
show that most clusters have marginal (1$\sigma$) individual
significance. If the 4--5\,Myr clusters represent the maximum possible
difference, it may be hard to achieve 3$\sigma$ significance for a
single cluster with $\sim$100 stars in the MB3 and MB4 bins. To
consider how large we expect these differences to be, we look at disk
evolution from a theoretical perspective.

To make an initial exploration of predictions for stellar mass
dependent disk dispersal, we consider two plausible disk evolution
models. To evaluate the observational signature of the evolution of
solid material in the disk, we first consider the possibility that
dust in the regions probed by the IRAC wavebands evolves more rapidly
for more massive stars. To illustrate observational diagnostics
derived from the global evolution of the disk, we then examine a
photoevaporation model, where a larger ionising flux causes disks
around more massive stars to disperse earlier. Although we do not
attempt to model the observational results in detail, we show that a
simple photoevaporation model reproduces the evolution of MB3 and MB4
disk fractions much better than our null hypothesis.

\subsection{Grain growth}\label{sec:growth}

Dust signatures decline due to removal of small grains. These grains
may be physically removed, or grow to unobservable sizes. Here, we
take grain growth to be that sufficient to remove any IR excess. The
timescale for growth depends on the orbital period. With the
simplifying assumption that a waveband traces dust at a single
temperature and radial distance, observations at fixed wavelength
probe different radial distances for different stellar
luminosities. Therefore, dust around different stars will be in an
earlier or later stage of growth. That is, for some stars there will
have been little growth and the dust is still observable and for
others it may have grown to invisible sizes.

To show how different stars observed with the same instrument can have
different grain growth timescales, we consider how the orbital period
changes with stellar mass at fixed disk temperature. The distance $a$
from a star of luminosity $L_\star$ to remain at a fixed temperature
$T$ is $a \propto L_\star^{-1/(4x)}$, where $T \propto a^{-x}$. For
PMS stars $L_\star \propto M_\star^2$, so $a \propto M_\star^{-1/2x}$
(i.e. a waveband probes greater distances around lower mass
stars). Because the period $P^2 \propto a^3/M_\star$ and $x \sim 3/4$
in the inner, less flared part of the disk
\citep[e.g.][]{1987ApJ...312..788A,1987ApJ...323..714K}, the relation
for period with stellar mass is
\begin{equation}
P \propto M_\star^{ \frac{3}{4x} - \frac{1}{2} } \sim \sqrt{ M_\star }
\end{equation}
That is, the greater luminosity of higher mass stars means the period
at fixed temperature increases with stellar mass.

However, the growth timescale $\tau_{\rm grow}$ also depends on the
surface density, which is generally thought to increase with stellar
mass \citep[based on mm observations,
e.g.][]{2000prpl.conf..559N,2005ApJ...631.1134A}. Thus,
\begin{equation}
\tau_{\rm grow} \propto P/\sigma \propto \frac{1}{ \sqrt{M_\star} }
\end{equation}
indicates a decreasing growth timescale with increasing stellar mass:
dust may disappear around higher mass stars more rapidly due to fast
grain growth timescales. This theory provides a qualitative
explanation for the much lower IR excesses found for early-type stars
that have disks.

While faster growth may provide some of the observed differential
evolution in disk fractions, it does not offer any explanation of why
the accretion signature also drops earlier for higher mass
stars. Indeed, if the presence of gas allows repeated fragmentation
and hinders growth \citep{2008A&A...486..597J}, then removal of the
gas may be more important in setting observational signatures of both
gas and dust. The leading theory for dispersal of the gaseous disk is
photoevaporation, to which we now turn.

\subsection{Photoevaporation}\label{sec:pe}

\begin{figure*}
\plottwo{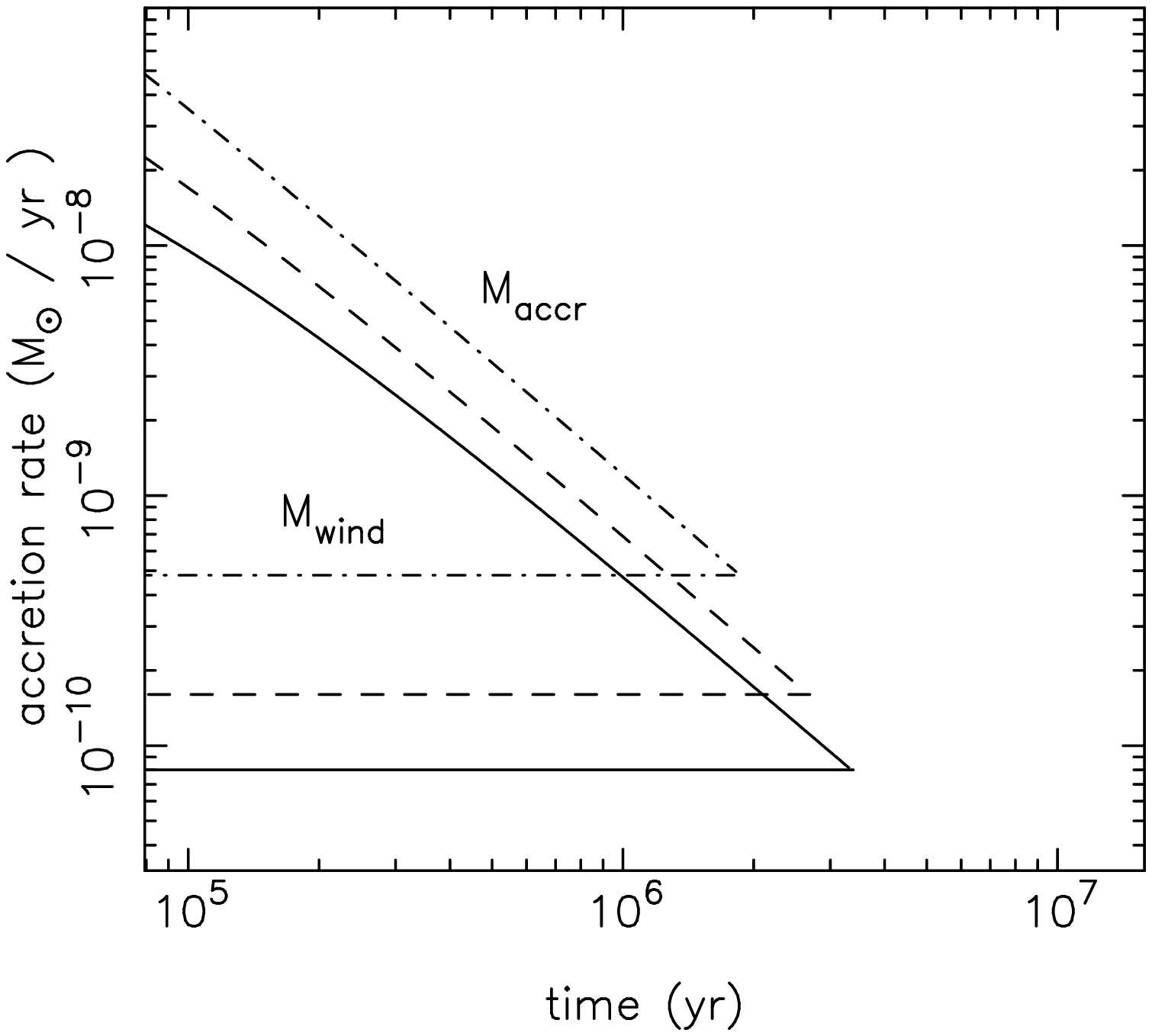}{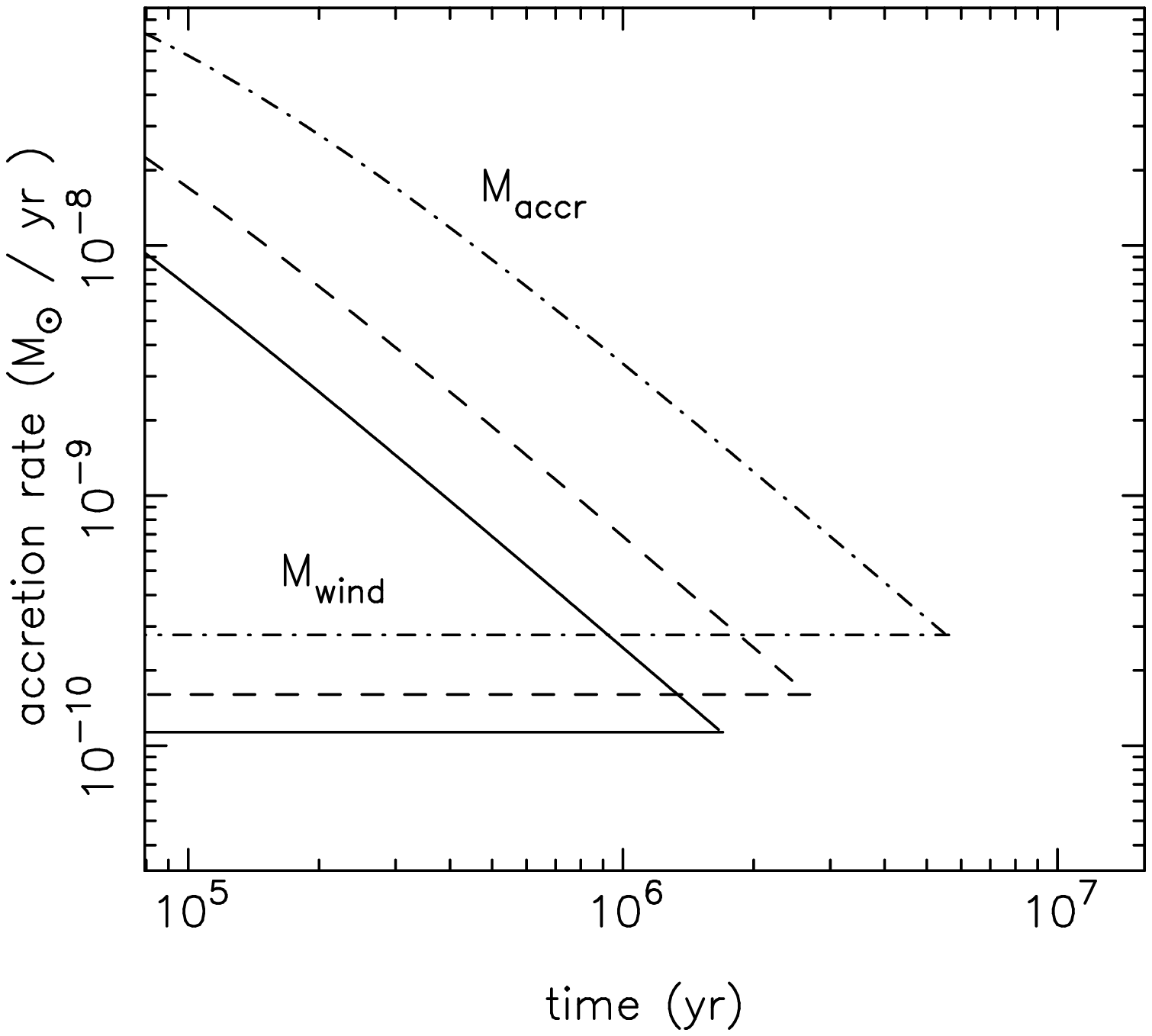}
\caption{Evolution of accretion for different choices of viscous
  evolution timescale: $P_\nu = -1$, $P_{\rm w} = 1$ (left), and
  $P_\nu = 1$, $P_{\rm w} = 1/2$ (right). Diagonal lines show
  accretion rates for 0.5 (solid), 1 (dashed), and 3\,$M_\odot$
  (dot-dashed), and horizontal lines show the wind loss rate for the
  same stellar masses. Disks are dispersed (and lines terminate) when
  the accretion rate drops below the wind loss rate. Thus, the disk
  lifetime decreases (left) or increases (right) with increasing
  stellar mass. }\label{fig:mdot}
\end{figure*}

Viscous evolution and photoevaporation work together to provide a
likely mechanism for removing the gaseous disks around young stars
\citep[e.g.][]{2001MNRAS.328..485C}. As the disk evolves and angular
momentum moves outward, gas moves inward and is accreted onto the
central star. At the same time, a wind of gas ionised by stellar
radiation removes material from the outer disk, beyond the ``critical
radius'' where the sound speed in the ionised gas is larger than the
orbital velocity \citep[e.g.][]{2000prpl.conf..401H}. When the
accretion rate through the inner disk onto the star drops below the
wind loss rate, the inner disk becomes depleted and is rapidly
accreted onto the star. With the inner disk removed, stellar radiation
rapidly photoevaporates the remaining gas \citep{2006MNRAS.369..229A}.

\subsubsection{A simple model}

To explore how this process impacts the time variation of disk
fractions in young clusters, we use a simple photoevaporation
model. This model captures the important aspects of the process
described above, with a prescription for the time evolution of the
accretion rate. When the accretion rate drops below the (fixed) wind
loss rate, the disk is considered dispersed
\citep{2006ApJ...639L..83A}.

For a viscous disk, the time dependent accretion rate is
\begin{equation}\label{eq:mdot}
\dot{M}_{\rm accr} = \frac{ M_{\rm disk}(0) }{ 2 t_\nu } \tau_\nu^{-3/2}\
\end{equation}
where $M_{\rm disk}(0)$ is the initial disk mass and $\tau_\nu =
t/t_\nu + 1$ is a dimensionless time. The viscous timescale is defined
at the scale radius $R_0$ by $t_\nu = R_0^2 / ( 3 \nu_0 )$, where the
viscosity is $\nu_0$. Initially, 1/$e$ of the disk mass lies outside
$R_0$
\citep[e.g.][]{1974MNRAS.168..603L,1998ApJ...495..385H,2006MNRAS.369..229A,2006ApJ...639L..83A}. Here
we assume $\nu \propto R$.

The disk rapidly disperses when the accretion rate drops below the
wind mass loss rate
\begin{equation}
\dot{M}_{\rm wind} \sim 1.6 \times 10^{-10}
\left( \frac{ \Phi }{ 10^{41} {\rm s}^{-1}} \right)^{1/2}
M_\star^{1/2} \, \,
\frac{ M_\odot }{ {\rm yr} }
\end{equation}
where $\Phi$ is the number of ionising photons per second and
$M_\star$ is in units of $M_\odot$
\citep{1994ApJ...428..654H,2004ApJ...607..890F}.

Applying this model to a range of stellar masses requires specifying
disk properties as a function of stellar mass. Specifically, this
variation needs to be defined for the viscous timescale and wind loss
rate. Assuming $t < t_\nu$, a typical disk temperature profile and
$M_{\rm disk} \propto M_\star$, \citet{2006ApJ...639L..83A} suggest
$t_\nu \propto M_\star^{-1}$ to match an apparent correlation of
accretion with stellar mass $\dot{M}_{\rm accr} \propto M_\star^2$
\citep[e.g.][]{2005ApJ...625..906M,2006A&A...452..245N}. Physically,
their model implies that the disk scale radius decreases with
increasing stellar mass. For the ionising flux, the
\citet{2006ApJ...639L..83A} model uses $\dot{M}_{\rm wind} \propto
M_\star$. Though they did not consider it significant, their model
predicts a stellar mass dependent disk lifetime (their Fig. 2). Higher
mass stars drive more powerful winds and have shorter accretion
timescales; thus, photoevaporation shuts off accretion earlier for
more massive stars.

Alternatively, \citet{2006MNRAS.369..229A} considered a linearly
increasing scale radius with increasing stellar mass, but fixed $\Phi$
(thus $\dot{M}_{\rm wind} \propto \sqrt{M_\star}$). They fixed $t_\nu$
to scale with the orbital timescale at the disk scale radius and
therefore $t_\nu \propto M_\star$. In this case the disk lifetimes
increase nearly linearly with stellar mass.

Thus, there is little constraint on how the viscous timescale changes
with stellar mass. Choosing $t_\nu \propto M_\star^{-1}$ based on the
apparently strong positive correlation between accretion and stellar
mass may be unfounded. Accretion measurements suffer a stellar mass
dependent bias \citep{2006MNRAS.370L..10C} and a linear relation
($\dot{M}_{\rm accr} \propto M_\star$) may be a better choice ($t_\nu$
independent of $M_\star$). Another parameter that has an impact is the
disk scale radius. If the initial disk radius increases with stellar
mass, then the viscous timescale increases with stellar
mass. Observationally, how disk radius varies with stellar mass is
poorly constrained. To explore the possibilities of the model, we
adopt $t_\nu = 2 \times 10^{-4} M_\star^{P_\nu}$\,yr ($P_\nu$: viscous
power-law index).

Some constraints on how the wind loss rate changes with stellar mass
exist. If the ionising flux scales with the bolometric stellar
luminosity and $L_\star \propto M_\star^2$ for PMS stars, then
$\dot{M}_{\rm wind} \propto M_\star^{1.5}$. A better indication of how
the high energy flux varies comes from the X-ray flux, which varies as
$\Phi_{\rm x} \propto M_\star^{1.5}$ \citep{2007A&A...468..353G},
yielding $\dot{M}_{\rm wind} \propto M_\star^{1.25}$. In our model, we
allow for variation of the wind loss rate by adopting $\dot{M}_{\rm
  wind} = 1.6 \times 10^{-10} M_\star^{P_{\rm w}} \, M_\odot / {\rm
  yr}$ ($P_{\rm w}$: wind power-law index).

The two cases outlined above then have $P_\nu = -1$ and $P_{\rm w} =
1$ \citep{2006ApJ...639L..83A} and $P_\nu = 1$ and ${P_{\rm w}} =
1/2$ \citep{2006MNRAS.369..229A}. For a typical disk mass, we follow
\citet{2006ApJ...639L..83A} and set $M_{disk}(0) = 0.01\,M_\star$
\citep[e.g.][]{2000prpl.conf..559N,2005ApJ...631.1134A}.

Figure \ref{fig:mdot} shows the two examples of disk evolution for a
range of stellar masses using Equation (\ref{eq:mdot}). The disk is
dispersed at the point where the accretion rate drops below the wind
rate (where lines of the same type meet and terminate). The left panel
shows the model with $P_\nu = -1$ and $P_{\rm w} = 1$; the disk
lifetime decreases with increasing stellar mass. The right panel shows
the other model, with $P_\nu = 1$ and $P_{\rm w} = 1/2$; the disk
lifetime increases with stellar mass.

Of these two models, the left panel qualitatively explains the trend
observed in our cluster data. The fact that all 5 clusters older than
3\,Myr show a decreased disk fraction for higher mass stars argues
that higher mass stars lose their disks earlier and that the model
that reproduces this behaviour is more realistic. This qualitative
agreement suggests that either the viscous timescale does not increase
with stellar mass, or the wind loss rate increases relatively strongly
with stellar mass.

To understand how the two models produce opposite trends in disk
lifetime with stellar mass, we look at how the wind rate and viscous
timescale change with stellar mass in more detail. Simplifying
Equation (\ref{eq:mdot}) by assuming $t > \tau_\nu$, setting
$\dot{M}_{\rm accr} = \dot{M}_{\rm wind}$ and solving for $t$
(i.e. the epoch of disk dispersal) yields
\begin{equation}\label{eq:taunu}
\tau_{\rm disk} \propto \frac{ M_{\rm disk} \sqrt{ t_\nu } }
{ \dot{M}_{\rm wind} } \propto M_\star^{1 + P_\nu/2 - P_{\rm w}} \, .
\end{equation}
This relation shows how disk lifetime varies with the viscous
timescale and wind loss rate. Thus, the disk lifetime decreases with
increasing stellar mass whenever $P_{\rm w} > 1 + P_\nu/2$. The two
models in Figure \ref{fig:mdot} thus have $\tau_{\rm disk} \propto
M_\star^{-1/2}$ and $\tau_{\rm disk} \propto M_\star$.

For our purposes, Equation (\ref{eq:taunu}) shows that the difference
between the scaling of viscous timescale and wind loss rate with
stellar mass sets how the disk dispersal time varies with stellar
mass. Though there are many uncertainties associated with the many
model parameters, we proceed with two of our own ``example'' models to
illustrate the implications of this model. These models also yield
estimates to compare with our observational results.

The first model has $t_\nu \propto 1 / M_\star$ and $\dot{M}_{\rm
wind} \propto M_\star$ ($P_\nu = -1$, $P_{\rm w} = 1$). The disk
lifetime is therefore $\tau_{\rm disk} \propto M_\star^{-1/2}$, the
same as the left panel of Figure \ref{fig:mdot}. Other solutions to
$P_{\rm w} = 1.5 + P_\nu/2$ yield the same results, provided the disk
lifetime is longer than the viscous timescale (true for the stellar
mass bins MB3 and MB4). The second model has $P_\nu = 0$ and ${P_{\rm
w}} = 1.25$ ($\tau_{\rm disk} \propto M_\star^{-1/4}$, the same as
other solutions to $P_{\rm w} = 1.25 + {P_\nu}/2$). This weaker
dependence illustrates how strongly the results depend on changes to
$P_{\rm w}$ and $P_\nu$.

\subsubsection{Application to cluster disk fractions}

To extend the model to a cluster of stars, we assume that a range of
disk lifetimes at fixed stellar mass arises from the natural
dispersion in initial disk masses within a cluster
\citep[e.g.][]{2000prpl.conf..559N,2005ApJ...631.1134A}. For fixed
stellar mass (and $\dot{M}_{\rm wind}$), more massive disks last
longer than less massive ones due to higher accretion rates. We assume
disk masses are normally distributed in log space about $M_{\rm disk}
= 0.01\,M_\star$ \citep[used by][]{2006ApJ...639L..83A}. A factor of
10 change in disk mass represents a 3\,$\sigma$ variation. Therefore,
at a given time, a fraction of disks have yet to be dispersed, which
is the observable quantity we compare the model with.

\begin{figure}
\plotone{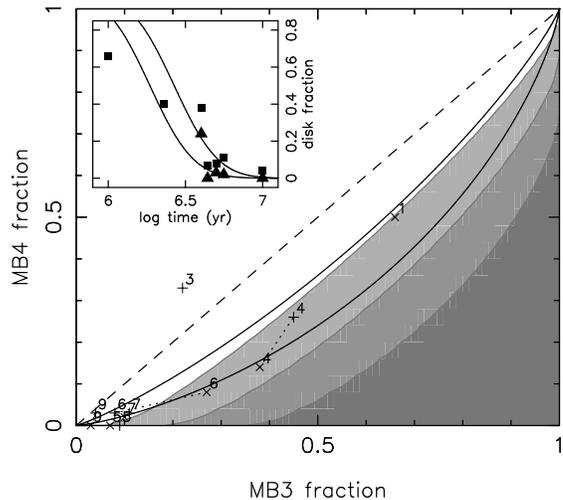}
\caption{Photoevaporation model with $\tau_{\rm disk} \propto
  M_\star^{-1/4}$ and $M_\star^{-1/2}$ compared to significance
  contours (solid lines). The dashed line shows our null hypothesis,
  photoevaporation independent of stellar mass. Data points are the
  same disk fractions from Figure \ref{fig:irac}, with error bars
  omitted for clarity. Filled regions show $>$1, 2, and 3\,$\sigma$
  significance from lightest to darkest. The inset shows the accretion
  model over time for $\tau_{\rm disk} \propto M_\star^{-1/2}$, with
  MB3 (squares) and MB4 (triangles) disk fractions overplotted (where
  the bin has $\ge$10 stars.)}\label{fig:accr}
\end{figure}

Figure \ref{fig:accr} shows the evolution of disk fraction for our two
cases of $\tau_{\rm disk} \propto M_\star^{-1/4}$ and $\tau_{\rm disk}
\propto M_\star^{-1/2}$, with stellar masses of 1 and
3\,$M_\odot$. The model stellar masses represent the MB3 and MB4 mass
bins in our cluster sample. The disk fraction loci start at the top
right of the figure, when all stars have disks. They then move down
and away from the line of equal disk fractions, as higher mass stars
lose their disks at a faster rate than lower mass ones. Finally, the
lines reach the lower left of the figure, when all stars have lost
their disks.

Though we do not attempt to fit a model, it is clear that $\tau_{\rm
 disk} \propto M_\star^{-1/2}$ is consistent with our observed
fractions. We compare the model and data in MB3 vs. MB4 space because
as shown by the Figure \ref{fig:tot-fr} and the Figure \ref{fig:accr}
inset, individual bin fractions vary widely. Attempting to fit the
time evolution of MB3 and MB4 would not yield useful or informative
parameters. We thus consider the relative difference between MB3 and
MB4 as disk fraction decreases, which appears to be a general property
of the $\gtrsim$3\,Myr clusters in our sample.

If our model is representative of the physical conditions in a
photoevaporating circumstellar disk, the relatively small differences
in observed disk fractions between mass bins may not be because there
is no signal (i.e. disk fraction is actually \emph{independent} of
stellar mass). It is because the signal is naturally weak and most
clusters have insufficient members for a significant result. Clusters
that fall in the times where the bulk of disks are being dispersed,
such as Tr 37, provide the best constraint how $\tau_{\rm disk}$
varies with $M_\star$. Though the 5\,Myr clusters have differences as
significant as Tr 37, the model disk fractions change little with
$P_\nu$ and $P_{\rm w}$ at these low disk fractions.

Calculating the perpendicular square residuals between the $\tau_{\rm
  disk} \propto M_\star^{-1/2}$ model and our data yields an extremely
good fit of $\chi^2 = 7.1$--2.1 (with the method as described for
testing our null hypothesis, \S \ref{sec:quant}). This result is much
smaller than for the null hypothesis and than expected given the
estimated errors. However, this comparison shows that a plausible
model reproduces the data much better than the null hypothesis.

To quantify the improvement in $\chi^2$, we use the Bayesian
Information Criterion \citep{Schwarz:1978p3785}, where $BIC = N \ln
\left( \chi^2 \right) + k \ln N$. While a model fit to $N$ points can
always be improved by adding parameters ($k$), the BIC tests whether
extra parameters lower the $\chi^2$ enough to be useful. Lower BIC
values are preferred; a difference of 2 between models indicates
positive evidence against the higher BIC value, a difference of 6
indicates strong evidence, and differences greater than 10 very strong
evidence.

For the two $\chi^2$ calculations of our null hypothesis ($N = 12$, $k
= 0$, $\chi^2 = 19.5$, and $N = 5$, $k = 0$, $\chi^2 = 23.4$), we find
$BIC = 36$ and 22. For the two $\chi^2$ calculations of our model ($N
= 12$, $k = 1$, $\chi^2 = 2.1$, and $N = 7$, $k = 1$, $\chi^2 = 7.1$),
we find $BIC = 11$ and 16. Thus, our model is \emph{significantly}
better at explaining the differences in disk fractions in MB3 and MB4
than the null hypothesis.

Given the degeneracy between the viscous timescale and wind loss rate,
we cannot put strong constraints on any parameters aside from
$\tau_{\rm disk}$. The data do not support models with higher wind
loss rates than observations of X-ray luminosity suggest, or stronger
than inverse dependence of the viscous timescale on stellar mass.

To estimate the significance of the observed and model disk fractions
in a more general way, we again use the binomial distribution. For
every point in MB3 and MB4 space where ${\rm MB3} > {\rm MB4}$
(i.e. below the dashed line in Figure \ref{fig:accr}), we estimate the
likelihood of observing those fractions when the intrinsic fraction is
a single value somewhere in between. This calculation is slightly
different than for finding individual errors. Instead of finding the
upper and lower intrinsic fractions that the observed fraction is
1\,$\sigma$ away from, we find the single intrinsic fraction with the
highest chance of observing the fractions in MB3 and MB4 (which are
different). Specifically, we use
\begin{equation}
\int_0^x B'(\epsilon;n_{\rm MB4},N_{\rm MB4}) d\epsilon =
\int_x^1 B'(\epsilon;n_{\rm MB3},N_{\rm MB3}) d\epsilon
\end{equation}
and solve for $x$. $B'$ is the probability distribution for intrinsic
fraction $\epsilon$, with sample size $N$ and observed number of disks
$n$ \citep[see][]{2003ApJ...586..512B}. The integrated area gives the
likelihood of measuring MB3 and MB4 from sampling stars with an
intrinsic disk fraction $x$.

To make this significance estimate, we need to choose properties of a
``typical'' cluster. A standard IMF
\citep[e.g.][]{2001MNRAS.322..231K} suggests there should be roughly
three times as many MB3 stars as MB4 stars. Table \ref{tab:data-mass}
shows a wide range in relative and overall numbers. Therefore we
choose the expected IMF ratio and use 100 stars (75 lower mass, 25
higher mass) over the two mass bins. This ``typical'' cluster
therefore has $\sim$1000 stars between 0.1--7\,$M_\odot$.  With the
caveat that only one of our cluster samples resembles this typical
cluster by number and IMF in MB3 and MB4 (the IR fraction for Tr 37),
this model allows us to roughly map out the entire disk fraction space
MB3 and MB4 may occupy.

The contours in Figure \ref{fig:accr} show the estimated significance
of differences in the two bins. The regions are $>$1, 2, and
3\,$\sigma$ from light to dark. Comparing the model to the contours,
the $\tau_{\rm disk} \propto M_\star^{-1/4}$ model suggests nearly
1\,$\sigma$ confidence is typical for samples with 100 stars in the
MB3 and MB4 bins. For $\tau_{\rm disk} \propto M_\star^{-1/2}$, the
difference in disk fractions generally sits within the 1\,$\sigma$
region. Thus, the $t_{\rm disk} \propto M_\star^{-1/2}$ model is
1--2\,$\sigma$ significant for this assumed typical cluster.

\subsubsection{Future observations}

\begin{figure}
\plotone{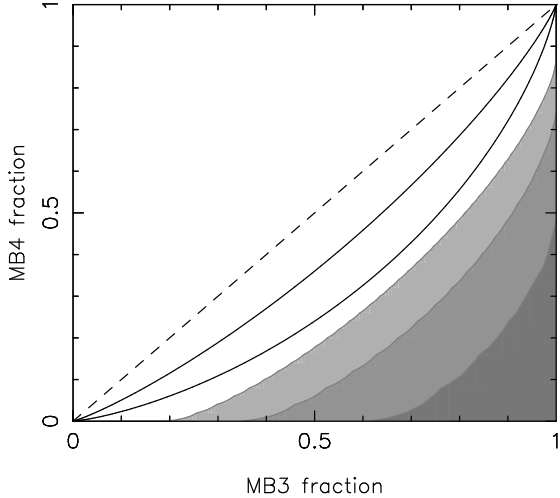}
\caption{Same as Figure \ref{fig:accr}, but contours show the number
  of stars needed for a 3\,$\sigma$ result: $<$48, 100, and 200 stars
  from darkest to lightest.}\label{fig:bsn}
\end{figure}

To illustrate how the significance contours in Figure \ref{fig:accr}
change with cluster size, we repeat the calculation for one larger,
and one smaller cluster. Figure \ref{fig:bsn} shows the number of
stars needed for a 3\,$\sigma$ result over the possible range of disk
fractions, with 48, 100, and 200 stars in MB3 and MB4. Overplotted are
evolutionary lines from the model. The Figure shows that $\gtrsim$200
stars are needed in the MB3 and MB4 bins for a 3\,$\sigma$ result if
$\tau_{\rm disk} \propto M_\star^{-1/2}$ and clusters show similar
results to the data and our model. For clusters following a typical
IMF, this requirement means a total of $\sim$2000 stars between 0.1
and 7\,$M_\odot$. However, it is important to remember that this is a
general estimate based on a typical IMF. The accretion fraction of Tr
37 shows that doubling the number of stars in MB4, with a similar
number in MB3 strongly increases the significance of the difference
(Tables \ref{tab:data-mass} and \ref{tab:fet}).

Though few young clusters with thousands of stars have been studied to
date, multi-object spectrographs such as Hectospec on the Multiple
Mirror Telescope (MMT) and 2df/AAOmega on the Anglo-Australian
Telescope (AAT), make spectroscopy of this many objects
possible. Clusters such as the Orion Nebula Cluster and h and $\chi$
Persei show that the desired numbers are obtainable. Despite being
$\sim$13\,Myr old, h and $\chi$ Persei shows evidence for stellar mass
dependent disk dispersal \citep{2007ApJ...659..599C}. A particularly
promising cluster is NGC 2264, which may have as many as 1000 stars,
and at 3\,Myr old is in a favourable age range for stellar mass
dependent disk dispersal. For other regions, such as NGC 2362 and
Orion OB1bc, additional work to obtain samples complete over the
widest possible range of spectral types will be beneficial.

A complementary way forward is to increase the number of clusters
studied. Here, we find higher mass stars in 5/5 clusters older than
$\sim$3\,Myr lose their disks faster than lower mass stars. If
additional clusters continue to show the same behaviour, the increased
numbers will strengthen this result. However, the significance of
individual clusters needs to be $\gtrsim$1\,$\sigma$ for the level at
which we reject the null hypothesis to increase. Otherwise the
$\chi^2$ value will increase by about the same amount as is required
by the extra degree of freedom for fixed confidence.

Thus, to show more strongly that stellar mass dependent disk dispersal
is a general result of disk evolution, obtaining the most complete
cluster samples possible is needed. Because most clusters will not
reach more than $\sim$2\,$\sigma$ significance, using larger samples
of clusters will also be important. The clusters most useful for
constraining disk dispersal models will be 3--5\,Myr old.

\section{Effects on planet formation}\label{sec:pf}

Our results in \S3 and \S4 suggest that the observed and predicted
signals for stellar mass dependent disk dispersal are real. While
additional clusters and larger samples of stars in 4--5\,Myr old
clusters may yield more significant results, there may be other
observable signatures of stellar mass dependent disk dispersal.

The orbits of giant planets provide a plausible test of this theory.
Gas giants require a substantial gas disk to form. Observations of
giant planets close to their parent stars suggest that planets migrate
inward after they form \citep[e.g.][]{1996Natur.380..606L}. Because
tidal torques between the planet and the disk are the most likely
migration mechanism, the masses and orbits of gas giants are plausibly
linked to the mass and lifetime of the disk. If these correlate with
stellar mass (\S4), the mass and orbits of gas giant planets may
correlate with the stellar mass.

\begin{figure}
\plotone{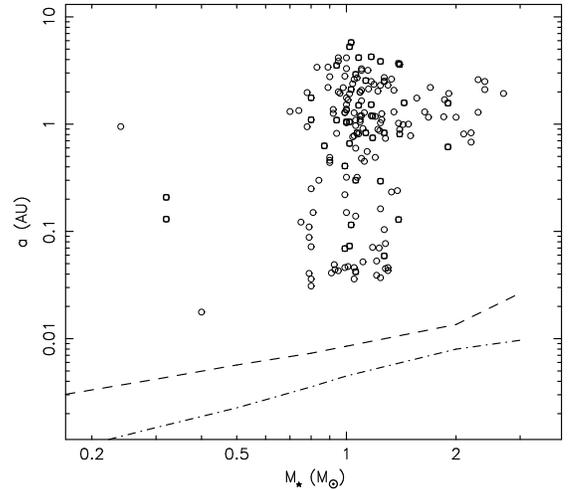}
\caption{Observed semi-major axis distribution of extra-Solar planets
  discovered by RV vs. host mass. Only planets with RV $> 30$\,m
  s$^{-1}$ are shown, roughly the signal from a 2\,$M_{\rm Jup}$
  planet at 1\,AU orbiting a 2\,$M_\odot$ star. Stellar radii from
  \citet{2000A&A...358..593S} tracks at $3 \times 10^6 / M_\star$
  (dashed line) and on the main-sequence (dot-dashed line) are also
  shown. There appears to be a real outward trend in semi-major axis
  for planets with host masses greater than
  $\sim$1\,$M_\odot$.}\label{fig:exopl}
\end{figure}

Observations of gas giants are starting to provide the data to guide
models. Figure \ref{fig:exopl} shows the semi-major axes of known
RV-discovered exoplanets as a function of stellar mass.\footnote{From
  http://exoplanet.eu , where $a$ and $M_\star$ known.}
%\footnote{From \href{http://exoplanet.eu}{exoplanet.eu}
To put these planets on an even footing, we only show planets with an
RV signal greater than 30\,m s$^{-1}$. This signal is about that
expected from a 2\,$M_{\rm Jup}$ planet at 1\,AU from a 2\,$M_\odot$
star and typical of planets orbiting $\gtrsim$Solar-mass
stars. Because the closest orbits are easiest to detect, the innermost
orbits at each stellar mass are probably close to the real
limits---interior planets of the same mass are easier to discover by
RV. These data suggest a small increase in the minimum separation
$a_{min}$ from 0.3\,$M_\odot$ ($a_{min} \approx 0.02$\,AU) to
1.6\,$M_\odot$ ($a_{min} \approx 0.03$\,AU) followed by a large jump
to $a_{min} \approx 0.61$\,AU at $\gtrsim$1.4\,$M_\odot$ \citep[see
also][]{2007ApJ...665..785J,2008PASJ...60..539S,2008arXiv0807.0268S,2008arXiv0810.1710N}. The
trend in $a_{min}$ with stellar mass is probably not a selection
effect.

Post-main sequence stellar evolution probably does not cause the
observed trend in $a_{min}$ with stellar mass. Current radial velocity
techniques are unable to achieve high accuracy for main sequence
A-type stars. Thus, to discover planets around intermediate mass
stars, radial surveys observe cooler, evolved objects. These stars
have larger radii than on the main-sequence, $\sim$5\,$R_\odot$ for a
2\,$M_\odot$ subgiant. If larger subgiant stars engulf close planets
or if tidal interactions cause close planets to spiral into the star,
massive subgiants would have many fewer close-in giant planets
\citep[e.g.][]{1996ApJ...470.1187R,2007ApJ...665..785J}.  However,
\citet{2007ApJ...665..785J} suggest that (i) only post--helium-flash
clump giants may have lost planets due to increased radii and (ii)
engulfing very close planets is not solely responsible for the lack of
short period planets around subgiants and K giants. This conclusion is
supported by numerical simulations \citep{2008PASJ...60..539S}.

If the step in $a_{min}$ at 1.6\,$M_\odot$ is not due to
post--main-sequence stellar evolution, then it is probably a signature
of the planet formation process. Here, we consider two ways to link
the orbits of gas giant planets with mechanisms of disk dispersal. In
the photoevaporation model, more rapid disk dispersal for intermediate
mass stars leads to a shorter time for gas giants to migrate closer to
their host stars. Thus, gas giants around more massive stars may have
larger orbits. In any model of disk dispersal, intermediate (low) mass
stars reach the main sequence before (after) the disk disperses. Thus,
the radius of a PMS star at the epoch of disk dispersal may set the
closest orbit for a gas giant planet.

\subsection{Migration}

To explore links between migration and disk dispersal, we consider
their relative timescales \citep[see
also][]{2004ApJ...616..567I,2007ApJ...660..845B}. Our analysis of the
cluster data shows that the disk dispersal timescale may decrease 
weakly with stellar mass.  For the photoevaporation model, 
$\tau_{\rm disk} \sim 3 \times 10^6 / M_\star^{1/2}$. If the migration 
timescale increases with (or is independent of) stellar mass, then the 
migration timescale will exceed the disk dispersal timescale at some 
stellar mass. For stars more massive than this limit, planets will 
remain on orbits near where they formed.

Planets undergoing type II migration are locked to the disk and move 
inward on the viscous timescale. A typical estimate of the migration
timescale $\tau_{\rm mig}$ is
\begin{equation}\label{eq:tmig}
  \frac{1}{\tau_{\rm mig}} = \frac{1}{a} \frac{da}{dt}
  \sim 1.5 \alpha h^2 \Omega \, ,
\end{equation} where $\alpha \sim 10^{-4}$ is a scaling parameter, $h
\sim 0.05$ is the disk aspect ratio, and $\Omega$ the orbital
frequency \citep{2002A&A...385..647D}.\footnote{Equation~(\ref{eq:tmig})
  applies when the planet mass is less than the local disk mass
  \citep[e.g.][]{1995MNRAS.277..758S}. The migration rate decreases as
  the planet begins to dominate. We reserve this detail for a more
  thorough model of formation and migration.} If the distance where
planets originate varies linearly with stellar mass \citep[e.g.][]{2008ApJ...682.1264K,2008arXiv0806.1521K} and
if $\alpha$ and $h$ are constant, the migration timescale is
$\tau_{\rm mig} \sim 1 \times 10^6 M_\star$\,yr.  

Setting $\tau_{\rm disk} = \tau_{\rm mig}$ yields $M_\star = 2$. Thus,
with our adopted disk dispersal and migration timescales, planets
around stars greater than $\sim$2\,$M_\odot$ should have their
migration halted by disk dispersal. Though this estimate is similar to
the observed transition mass between close and more distant orbits, it
is highly uncertain. For example, because type II migration is linked
to the viscous evolution of the disk, how migration changes with
stellar mass is uncertain (as discussed in \S \ref{sec:pe}).
Exploring this picture in detail requires a detailed model that
considers concurrent migration and disk dispersal as a giant planet
grows.

This picture has two other implications for giant planet formation.
In the core accretion model for gas giant planet formation,
protoplanets grow more slowly at larger distances from their host
star. Thus, the disk lifetime effectively sets an outer limit to where
gas giants form \citep[e.g.][]{2008ApJ...673..502K}. If the disk
lifetime depends on stellar mass, this outer limit is closer than
predicted by models with a fixed disk lifetime
\citep[e.g.][]{2005ApJ...626.1045I,2008ApJ...673..502K}.

Differential disk dispersal rates may also affect giant planet
frequency.  Current observations suggest an increasing frequency of
gas giants around more massive stars
\citep{2006PASP..118.1685B,2007ApJ...670..833J}. In planet formation
theory, this frequency is set by two competing effects: (i) shorter
disk lifetimes for more massive stars reduce the likelihood of forming
giant planets and (ii) higher disk masses for more massive mass stars
increase the probability of gas giant planet formation.  As for the
implications for migration, understanding how these effects affect gas
giant formation requires more detailed models that are beyond the
scope of this paper.

\subsection{Pre--main-sequence contraction}

To conclude this section, we consider whether PMS stellar evolution
can affect the orbits of close in giant planets. As young stars
approach the main sequence, they contract. The PMS contraction, which
is ongoing during giant planet formation and disk dispersal, may
therefore affect the innermost orbits that migrating planets may
reach.

The inner edge of a circumstellar disk (the truncation radius) is a
function of the stellar radius \citep[e.g.][]{2007prpl.conf..479B}. If
planets cannot migrate interior to this edge
\citep[e.g.][]{1996Natur.380..606L}, then the stellar radius at the
epoch of disk dispersal sets the innermost possible orbit \citep[aside
from later movement due to tidal evolution, which also depends on
stellar parameters
e.g.][]{1996ApJ...470.1187R,2008arXiv0801.0716J}. The larger radii of
more massive stars and their shorter disk dispersal timescales,
prevent planets from reaching closer orbits during their PMS phase.

Because the radius of a PMS star changes with time, we can use our
disk dispersal timescale to estimate the stellar radius when the disk
is removed. Observationally, 5--10 stellar radii represents roughly
the closest orbit a planet can reach by migration, which must occur
while the disk is still present. The dashed line in Figure
\ref{fig:exopl} shows how the stellar radius at $\tau_{\rm disk} = 3
\times10^6 / M_\star^{1/2}$\,yr varies with stellar mass using the
\citet{2000A&A...358..593S} PMS tracks. The PMS stellar radii shows a
weak trend with stellar mass, similar to the innermost orbits. Thus,
the innermost orbits of exoplanets may be set by the radius of their
PMS host star, through the stars influence on the inner edge of the
circumstellar disk.

This suggestion is uncertain for many reasons. For example, the disk
inner radius may not be a linear function of the stellar radius and
orbits evolve through tidal interaction with the host star after
reaching close orbits. Future discoveries of exoplanets orbiting
low-mass stars will fill in the left side of Figure \ref{fig:exopl},
giving a better idea of how the innermost orbit changes with stellar
mass.

\subsection{Alternatives and future work}

The diversity of planet formation models means there are alternative
theories that may explain the observed orbits of intermediate mass
stars. These theories suggest the trend is either a formation
signature, or a result of later stellar evolution. These theories make
testable predictions, that will be judged based on future
observations.

\citet{2008arXiv0806.1521K} suggest that the larger orbits may be a
signature of the ``dead zone'' in a layered disk accretion model
\citep{1996ApJ...457..355G}. The inner edge of the dead zone (whose
distance varies roughly linearly with stellar mass in their model)
acts as a ``trap'' for both planets
\citep[e.g.][]{2006ApJ...642..478M} and their building
blocks. However, if giant planet orbits were purely a result of the
dead zone distance, one might expect a roughly linear dependence of
planet semi-major axes on the mass of their hosts, rather than the
step that appears in Figure \ref{fig:exopl}.

One first step toward understanding the origin of the larger orbits
for intermediate mass stars is to verify that the trend is
real. Though difficult, if close-in planets can be found orbiting
intermediate mass stars before they leave the main-sequence, then the
engulfment scenario may be ruled out. Further work using planet
formation models
\citep[e.g.][]{2007ApJ...660..845B,2008ApJ...673..502K} can study how
changes in disk evolution affect the observable outcomes of planet
formation. The continued discovery of planets around low and
intermediate-mass stars will provide further constraints on the final
outcomes these models must produce.

As we have shown, planet formation models provide a link between disk
evolution and observed exoplanet distributions. Thus, the inclusion of
differential disk lifetimes in these models can attempt to understand
both how planets form and how the disks they form in evolve.

\section{Summary}\label{sec:summary}

Our results suggest a stellar mass dependent timescale for the
dispersal of circumstellar disks around young stars. Intermediate-mass
stars tend to lose their disks earlier than Solar-mass stars. All
clusters in our study older than $\sim$3\,Myr show this tend. We
reject the null hypothesis---that Solar and intermediate-mass stars
lose their disks at the same rate---with 95--99.9\% confidence. For
each cluster, higher mass stars lose their disks earlier than their
Solar mass counterparts with a significance of roughly
1\,$\sigma$. For low mass stars, there is a clear disagreement in disk
dispersal timescales derived from accretion and dust signatures. This
discrepancy may be partly due to an increased occurrence of transition
disks among lower-mass stars.

By considering how the timescale for grain growth varies with stellar
mass for a fixed disk temperature, we show that the dust around higher
mass stars may appear to be more evolved than lower mass stars. This
model provides a possible explanation for the lower IR excesses
observed for early-type stars with disks, as compared to later-type
stars.

Our analysis of a reasonable photoevaporation model demonstrates that
the predicted signature of stellar mass dependent disk dispersal is
subtle. In this model, the largest differences in disk fractions are
expected when the bulk of cluster stars are dispersing their
disks. Observations suggest that this dispersal occurs between
4--5\,Myr. At earlier (later) times, we expect small differences
because most (no) stars of all masses have disks.

Though it is hard to rule out the increasing multiplicity fraction
with stellar mass as an alternative mechanism, the lack of any
observed differences in disk evolution in multiple systems argues for
photoevaporation as the likely mechanism. As noted in
\ref{sec:binarity}, the effects of binary companions are complex and
largely unknown. Given that around half of stellar systems may be
binaries or multiples in the primary mass range where planets are
routinely discovered \citep[e.g.][]{1991A&A...248..485D}, and that
planets are known to exist in binary systems
\citep[e.g.][]{2003ApJ...599.1383H}, the effects of binaries merit
further study. To make progress in this area requires larger samples
of objects with known multiplicity. The most useful samples will
contain objects with $\lesssim$10\,AU separations, where the effects
on disk evolution as observed by accretion and hot-dust signatures are
thought to be strongest. Obtaining samples with a wide range of
primary masses will be essential to understand the effects (if any) on
stellar mass dependent disk dispersal.

Stellar mass dependent disk dispersal may have consequences for
extra-Solar planets observed around main-sequence and older stars. 
Current observations find a step in planet semi-major axes for
stars more massive than 1.6\,$M_\odot$. This feature may be caused by
a shorter disk dispersal timescale for more massive stars. Giant
planets forming around these stars have less time to migrate and
remain on orbits near where they form. More planet detections over a
range of stellar masses will test the reality of the apparent step in
$a_{min}$ at 1.6\,$M_\odot$ and allow better tests of models for
migration and disk dispersal. Studying tidal decay in more detail will
also indicate the level at which stellar evolution affects these
orbits.

The causes and effects of stellar mass dependent disk dispersal are
many-fold and complex and progress can be made in several
directions. For young stars, high resolution spectroscopy and direct
detection of the H$_2$ component of circumstellar disks will yield
better knowledge of how the gaseous component evolves. For low-mass stars, consideration of the local environment may show
that disk dispersal depends on proximity to luminous O stars. A
greater knowledge of multiplicity on an individual level will allow
further studies of how companions may affect disk evolution.

Testing our main result in more detail requires larger samples. To
reject the null hypothesis with greater confidence requires both more
clusters and a high level of completeness for new and known
clusters. Because the error drops roughly as $1 / \sqrt{N}$, large
increases in significance for well studied clusters will be
difficult. However, clusters such as NGC 2362 and Orion OB1bc and OB1a
have many more Solar and intermediate-mass stars that need their
circumstellar environments characterised, so will benefit from further
study.

Additional clusters are also needed. Clusters with ages of $\sim$ 4--5
Myr---when most stars lose their disks---provide the most sensitive
measure of the dispersal time as a function of stellar mass. With many
intermediate-mass stars and at $\sim$3\,Myr, NGC 2264 is probably the
best example of a rich cluster with unpublished Spitzer IRAC data. It
will be interesting to see whether this cluster shows results similar
to the slightly older Tr 37, or the slightly younger IC 348. Obtaining
spectroscopy of many objects in rich clusters and associations is made
possible with multi-object spectrographs such as Hectospec and
2df/AAOmega.

Because there will be only a few clusters with the several thousand
stars required for high significance, progress will come from
increases in both cluster numbers and the best possible level of
completeness for all clusters.

\acknowledgements

We thank Cathie Clarke and Charles Jenkins for helpful conversations,
and the referee for a thorough review. GK thanks Mark Wyatt and the
Cambridge IoA, and the Harvard-CfA, where part of this study was
carried out. This research was supported by an Australian Postgraduate
Award and an ANU Vice-Chancellor's Travel Grant (GK), and the {\it
  NASA Astrophysics Theory Program} through grant NAG5-13278 and the
\emph{TPF Foundation Science Program} though grant NNG06GH25G
(SK). This research has made use of the SIMBAD database and the VizieR
catalogue access tool, both operated at CDS, Strasbourg, France. It
also makes use of data products from the Two Micron All Sky Survey,
which is a joint project of the University of Massachusetts and the
Infrared Processing and Analysis Center/California Institute of
Technology, funded by the National Aeronautics and Space
Administration and the National Science Foundation.

\bibliography{ref.bib,extras.bib} \bibliographystyle{astroads}

\clearpage

\begin{deluxetable*}{lcccc}
\tablecolumns{5} \tablecaption{Disk fractions for single and multiple stars\label{tab:binary}.}
  \tablehead{ \colhead{} & \multicolumn{2}{c}{Single} & \multicolumn{2}{c}{Binaries} \\   \colhead{Name} & \colhead{\ha} & \colhead{IR} & \colhead{\ha} & \colhead{IR} }
\startdata
Taurus & 40/55=72\% & 42/55=76\% & 33/50=66\% & 37/50=74\% \\
Upper Sco &  2/49= 4\% &  2/53= 3\% &  2/37= 5\% &  1/42= 2\% \\
\enddata
\end{deluxetable*}

\begin{deluxetable*}{lccccccccc}
\tabletypesize{\footnotesize}
\tablecolumns{10} \tablecaption{Cluster disk fractions binned by spectral type\label{tab:data-spty}.}
  \tablehead{ \colhead{} & \colhead{Age} & \multicolumn{2}{c}{M} & \multicolumn{2}{c}{K} & \multicolumn{2}{c}{FG} & \multicolumn{2}{c}{BA} \\   \colhead{Name} & \colhead{(Myr)} & \colhead{\ha} & \colhead{IR} & \colhead{\ha} & \colhead{IR} & \colhead{\ha} & \colhead{IR} & \colhead{\ha} & \colhead{IR} }
\startdata
1 Taurus & 1 & 42/55=76\% & 50/67=75\% & 24/40=60\% & 24/40=60\% & 2/4=50\% & 1/2=50\% & 2/3=67\% & 2/2=100\% \\
2 Cha I & 2 & 37/107=35\% & 46/87=53\% & 13/23=57\% & 9/13=69\% & 2/2=100\% & 1/2=50\% & 1/2=50\% & \ldots \\
3 IC 348 & 2.3 & 46/173=27\% & 102/233=44\% & 7/12=58\% & 8/27=30\% & 3/3=100\% & 3/11=27\% & \ldots & 2/7=29\% \\
4 Tr 37 & 4 & 21/53=40\% & 28/48=58\% & 31/69=45\% & 30/59=51\% & 4/25=16\% & 3/9=33\% & 2/35=6\% & 1/13=8\% \\
5 NGC 2362 & 5 & 4/86=5\% & 7/65=11\% & 2/35=6\% & 2/35=6\% & 0/4=0\% & 0/5=0\% & 0/7=0\% & 0/7=0\% \\
6 OB1bc & 5 & 13/85=15\% & 4/31=13\% & 21/68=31\% & 1/11=9\% & 0/10=0\% & 0/8=0\% & 6/103=6\% & 3/103=3\% \\
7 Upper Sco & 5 & 14/183=8\% & 13/92=14\% & 2/29=7\% & 5/25=20\% & 1/31=3\% & 0/35=0\% & 4/71=6\% & 3/78=4\% \\
8 OB1a/25Ori & 8 & 3/68=4\% & 1/38=3\% & 4/23=17\% & 2/8=25\% & 1/1=100\% & \ldots & \ldots & \ldots \\
9 NGC 7160 & 10 & 0/12=0\% & 1/8=12\% & 1/21=5\% & 1/17=6\% & 0/62=0\% & 1/47=2\% & 0/35=0\% & 1/34=3\% \\
\enddata
\end{deluxetable*}

\begin{deluxetable*}{lccccccccc}
\tabletypesize{\footnotesize}
\tablecolumns{10} \tablecaption{Cluster disk fractions binned by mass\label{tab:data-mass}.}
  \tablehead{ \colhead{} & \colhead{Dist} & \multicolumn{2}{c}{MB1 (68\%)} & \multicolumn{2}{c}{MB2 (20\%)} & \multicolumn{2}{c}{MB3 (9\%)} & \multicolumn{2}{c}{MB4 (3\%)} \\   \colhead{Name} & \colhead{(pc)} & \colhead{\ha} & \colhead{IR} & \colhead{\ha} & \colhead{IR} & \colhead{\ha} & \colhead{IR} & \colhead{\ha} & \colhead{IR} }
\startdata
1 Taurus & 140 & 3/4=75\% & 7/13=54\% & 31/42=74\% & 34/45=76\% & 27/41=66\% & 27/41=66\% & 5/10=50\% & 5/7=71\% \\
2 Cha I & 165 & 13/52=25\% & 21/47=45\% & 19/39=49\% & 14/22=64\% & 11/20=55\% & 7/11=64\% & 4/6=67\% & 2/3=67\% \\
3 IC 348 & 320 & 21/96=22\% & 52/125=42\% & 12/41=29\% & 21/54=39\% & 6/16=38\% & 7/32=22\% & 4/4=100\% & 6/18=33\% \\
4 Tr 37 & 900 & \ldots & \ldots & 17/38=45\% & 23/33=70\% & 34/89=38\% & 34/75=45\% & 7/50=14\% & 5/19=26\% \\
5 NGC 2362 & 1500 & 1/7=14\% & 0/3=0\% & 2/67=3\% & 5/51=10\% & 3/43=7\% & 4/43=9\% & 0/14=0\% & 0/14=0\% \\
6 OB1bc & 440 & 0/4=0\% & 1/3=33\% & 11/72=15\% & 3/26=12\% & 21/77=27\% & 1/12=8\% & 7/93=8\% & 3/92=3\% \\
7 Upper Sco & 145 & 7/103=7\% & 9/50=18\% & 7/73=10\% & 5/43=12\% & 3/31=10\% & 3/28=11\% & 2/92=2\% & 3/96=3\% \\
8 OB1a/25Ori & 322 & 0/11=0\% & 0/7=0\% & 1/43=2\% & 1/29=3\% & 7/38=18\% & 2/10=20\% & \ldots & \ldots \\
9 NGC 7160 & 900 & \ldots & \ldots & 0/12=0\% & 1/8=12\% & 1/38=3\% & 1/27=4\% & 0/80=0\% & 2/71=3\% \\
\enddata
\end{deluxetable*}

\begin{deluxetable}{lccc}
\tablecolumns{4} \tablecaption{Fishers Exact Test for MB3 and MB4\label{tab:fet} (\%).}
\tablehead{ \colhead{Name} & \colhead{\ha} & \colhead{IR} & \colhead{Common} }
\startdata
1 Taurus & 28.23 (10) & 75.74 (7) & 68.04 (6) \\
2 Cha I & 83.48 (6) & 76.92 (3) & 100.00 (1) \\
3 IC 348 & 100.00 (4) & 88.83 (18) & 100.00 (3) \\
4 Tr 37 & 0.19 (50) & 10.58 (19) & 20.50 (17) \\
5 NGC 2362 & 42.18 (14) & 31.24 (14) & 39.83 (14) \\
6 OB1bc & 0.05 (77) & 39.23 (12) & 40.20 (12) \\
7 Upper Sco & 10.13 (31) & 12.79 (28) & 14.41 (19) \\
8 OB1a/25Ori & \ldots (0) & \ldots (0) & \ldots (0) \\
9 NGC 7160 & 32.20 (38) & 62.42 (27) & 28.13 (27) \\
\enddata
\end{deluxetable}

\end{document}